\documentclass[3p,times,review,12pt,showpacs,showkeys,pre,amsmath,amssymb,superscriptaddress]{elsarticle}

\usepackage{epsfig}
\usepackage{color}
\usepackage{nicefrac}
\usepackage{booktabs}
\usepackage{graphicx}
\usepackage{natbib}
\usepackage{amsmath}
\usepackage{hyperref}
\usepackage{amssymb}
\usepackage{amsthm, mathtools}

\usepackage{color}

\usepackage{float}
\usepackage{algorithm,algpseudocode, esint}

\usepackage[%
  font=small,
  labelfont=small,
  textfont=small,
  format=hang,
  format=plain,
  margin=0pt,
  width=0.8\textwidth,
]{caption}
\usepackage[list=true]{subcaption}

\graphicspath{{./}}

\begin{document}

\begin{frontmatter}


  \title{A numerical tool for the study of the hydrodynamic recovery of the Lattice Boltzmann Method \footnote{Postprint version of the article published on Computers \& Fluids 172 (2018) 241-250} }

  \author[UTOV,BUW]{Guillaume Tauzin\corref{cor}}
  \ead{guillaume.tauzin@roma2.infn.it}
  \author[UTOV]{Luca Biferale}
  \author[UTOV]{Mauro Sbragaglia}
  \author[TUE]{Abhineet Gupta}
  \author[TUE]{Federico Toschi}
  \author[BUW]{Andreas Bartel}
  \author[BUW]{Matthias Ehrhardt}
  \cortext[cor]{Corresponding author:}
  \address[UTOV]{Dipartimento di Fisica and INFN,
    Universit\`{a} di Roma ``Tor Vergata",
    Via della Ricerca Scientifica 1, 00133 Roma, Italy}
  \address[TUE]{Department of Applied Physics and
    Department of Mathematics and Computer Science,
    Eindhoven University of Technology,
    5612 AZ Eindhoven, Netherlands}
  \address[BUW]{Chair of Applied Mathematics and Numerical Analysis,
    Bergische Universit\"{a}t Wuppertal,
    Gau\ss{}strasse 20, 42119 Wuppertal, Germany}

\begin{abstract}
{\color{black} We investigate the hydrodynamic recovery of Lattice Boltzmann Method (LBM) by analyzing exact balance relations for energy and enstrophy derived from averaging the equations of motion on sub-volumes of different sizes.} In the context of 2D isotropic homogeneous turbulence, we first validate this approach on decaying turbulence by comparing the hydrodynamic recovery of an ensemble of LBM simulations against the one of an ensemble of Pseudo-Spectral (PS) simulations. We then conduct a benchmark of LBM simulations of forced turbulence with increasing Reynolds number by varying the input relaxation times of LBM. This approach can be extended to the study of implicit subgrid-scale (SGS) models, thus offering a promising route to quantify the implicit SGS models implied by existing stabilization techniques within the LBM framework.
\end{abstract}


\end{frontmatter}

\section{Introduction}
The simulation of turbulent flows pertains to a vast diversity of applications in engineering~\cite{Galperin2010}. The high Reynolds number associated with the phenomenon of turbulence requires solving a wide range of scales on a high resolution computational grid, making their Direct Numerical Simulation (DNS) typically out of reach~\cite{Pope2000, Davidson2015}. Large-Eddy Simulation (LES) is a workaround which allows a reduction of the number of degrees of freedom. LES is acknowledged in the engineering community as a cost-effective alternative to DNS~\cite{Pitsch2006, Wagner2007, Sullivan1994}. The principle of LES is to solve flow scales up to a cut-off and to filter the small scales out. As large scales and smaller scales are coupled, unresolved small scales need to be modeled using a so-called subgrid-scale (SGS) model. A large number of filtering techniques and SGS models have been proposed in the Navier-Stokes framework~\cite{Sagaut2002}.\\
The Lattice Boltzmann Method (LBM) is a meso-scale flow solver that has been gaining popularity because of its intrinsic scalability, as well as its ability to deal with multiple physics and complex boundary conditions~\cite{Succi2001, Wolf-Gladrow2000, Kruger2017}. The LBM equation describes the streaming and collision of distribution functions $f_{\ell}(\vec{x},t)$ on a lattice with a finite set of kinetic velocities $\vec{c}_{\ell},\,\, {\ell}=0 \ldots q-1$. The collision operator is popularly modeled by the Bhatnagar-Gross-Krook (BGK)~\cite{Bhatnagar1954} relaxation towards a local equilibrium with a dimensionless relaxation time $\tau$
\begin{equation}
\label{eq:LBM}
f_{\ell}(\vec{x}+\vec{c}_{\ell} \Delta t,t + \Delta t) - f_{\ell}(\vec{x},t)= -\frac{1}{\tau}\left[ f_{\ell}(\vec{x},t)-f^{eq}_{\ell}(\vec{x},t)\right]+F_{\ell}
\end{equation}
where $F_{\ell}$ is a suitable forcing term designed to reproduce a macroscopic forcing~\cite{Succi2001, Wolf-Gladrow2000, Kruger2017}. From a theoretical point of view, the use of a multi-scale Chapman-Enskog (CE) perturbative expansion allows to recover hydrodynamic equations. In brief, one expands the distribution function in a power-series: $f_{\ell}=f^{(eq)}_{\ell}+K_n f^{(1)}_{\ell}+K^2_n f^{(2)}_{\ell}+...$, where $K_n=\lambda/L \ll 1$ is the Knudsen number, giving the ratio between the particles mean free path $\lambda$ and the macroscopic scale $L$. Furthermore, space and time are rescaled, \textit{i.e.} $\vec{x}^{(1)}=K_n \vec{x}$, $t^{(1)}=K_n t$, $t^{(2)}=K_n^2 t$ by introducing separate time scales for the effect of advection ($t^{(1)}$) and dissipation ($t^{(2)}$)~\cite{Succi2001,Wolf-Gladrow2000}. Performing this procedure for a local equilibrium distribution chosen as (repeated indices are meant summed upon)

\begin{equation}
\label{eq:feq}
f^{eq}_{\ell} ( \vec{x}, \, t)
= f^{eq}_{\ell} \left( \rho (\vec{x}, \, t), \, \vec{u} (\vec{x}, \, t)  \right)
=  t_{\ell} \, \rho \left[ 1 + \frac{c_{\ell, \, i} u_i}{c^2_s} + \frac{\left( c_{\ell, \, i} u_i \right)^2}{2 c^4_s} - \frac{u_i u_i}{2 c^2_s} \right],
\end{equation}
where $t_{\ell}$ is a set of lattice-dependent weighting factors and $c_s$ the speed of sound in the lattice, one can recover the athermal weekly compressible Navier-Stokes hydrodynamic equations for the density field $\rho(\vec{x}, \, t)=\sum^{q-1}_{\ell=0} f_{\ell}(\vec{x}, \,t)$ and velocity field $\vec{u}(\vec{x}, \, t)=\sum^{q-1}_{\ell=0} f_i(\vec{x}, \, t) \, \vec{c}_{\ell}/\rho(\vec{x}, \,t)$

\begin{equation}\label{eq:mass_LBM}
\partial_t \rho + \partial_j (\rho u_j)=0 + \mathcal{O} (K_n^2)
\end{equation}
\begin{equation}\label{eq:N-S_LBM}
\partial_t \left( \rho u_i \right) + \partial_j \left( \rho u_i u_j \right)
= - \partial_i p + \partial_j \left( \rho \nu \left( \partial_j u_i + \partial_i u_j \right) \right) + F_i
+ \mathcal{O} (K_n^2) + \mathcal{O} (Ma^3).
\end{equation}
Beyond the higher order corrections in the Knudsen number, in the recovery of the momentum equations one usually neglects terms which are cubic in the velocity \cite{Viggen2009}, hence we find the term $\mathcal{O} (Ma^3)$, where the Mach number $Ma = \frac{U_{RMS}}{c_s}$ represents the ratio of the root mean square velocity $U_{RMS}$ to $c_s$. The term $p = c_s^2 \rho$ is the fluid pressure and the viscosity $\nu$ is linearly dependent on the relaxation time $\tau$ in \eqref{eq:N-S_LBM_nu} and vanishes as $\tau \rightarrow 0.5$:
\begin{equation}\label{eq:N-S_LBM_nu}
\nu=c_s^2 \left( \tau - \frac{1}{2} \right) \Delta t.
\end{equation}
The LBM community has been keenly proposing Navier-Stokes inspired LES techniques to combine the intrinsic scalability of LBM with turbulence SGS models. The majority of them are eddy viscosities models implemented by locally modifying the relaxation time $\tau$, i.e. assuming that Eq.~\eqref{eq:N-S_LBM_nu} holds and that an effective relaxation time $\tau_{\text{eff}} (\vec{x}, \, t)$ results in an effective viscosity $\nu_{\text{eff}} (\vec{x}, \, t)$~\cite{Filippova2001a, Dong2008a, Dong2008b, Chen2009}. Malaspinas \& Sagaut have shown that this method is only valid in the athermal weakly compressible limit and proposed a consistent eddy viscosity closure extension for compressible thermal flows~\cite{Malaspinas2012}. Instabilities of the LBM with a BGK collision operator (LBGK) arising for an input relaxation time $\tau_0 \rightarrow 0.5$, \textit{i.e.} for an input viscosity $\nu_0 \rightarrow 0$, along with the low $Ma$, which is required to remain in a good approximation of Navier-Stokes, significantly limit the range of Reynolds number reachable in practice. Some eddy viscosity methods have been shown to extend the range of stability to relaxation times $\tau_0 \rightarrow 0.5$, making it possible to simulate higher Reynolds number flows for a fixed grid resolution~\cite{Premnath2009}. Stabilization of LBGK has been linked to the existence of an underlying Lyapunov functional in the form of a discrete Boltzmann H-functional~\cite{Boghosian2001}. Karlin \textit{et al.}~\cite{Karlin1999} introduced the Entropic Lattice Boltzmann (ELBM): an LBGK ensuring the monotonicity of a convex H-functional commonly chosen as
\begin{equation}
H \left( \mathbf{f} \right) = \sum^{q-1}_{\ell=0} f_{\ell} \log \left( \frac{f_{\ell}}{t_{\ell}} \right),  \,\, {\bf f} = \left\{ f_{\ell} \right\}^{q-1}_{\ell=0}.
\end{equation}
To equip a LBGK with an H-theorem, ELBM implements a collisional process with an effective relaxation time $\tau_{\text{eff}} = \frac{2 \tau_0}{\alpha}$ to a local equilibrium distribution $\mathbf{f^{eq}}$ defined as the extremum of the H-functional under the constraints of mass and momentum conservation. The parameter $\alpha$ is calculated locally (in space and time) and has a non-linear dependency on the distribution functions $f_{\ell}$. While the result is an unconditionally stable LBGK for $\tau_0 \rightarrow 0.5$ ($\nu_0 \rightarrow 0$), we are also left with a side-effect effective viscosity $\nu_{\text{eff}}$. Unfortunately, the non-linear dependency of the effective relaxation time on the distribution functions does not allow this effective viscosity to be expressed in terms of macroscopic quantities and therefore the physics behind it remains hidden. In 2008, Malaspinas \textit{et al.}~\cite{Malaspinas2008} proposed an approximate formulation of the effective viscosity $\nu_{\text{eff}} (\vec{x}, \, t) = \nu_0 + \nu_t (\vec{x}, \, t)$ using CE expansion assuming $\alpha \approx 2$ ($\tau_{\text{eff}} \approx \tau_0$). The resulting turbulent viscosity $\nu_\text{t}$ is
\begin{equation}\label{eq:viscosity_malaspinas}
\nu_{\text{t}} = - \frac{c^{2}_{s}}{3} \tau_0^2 \Delta t^2
\frac{S_{\theta \kappa} S_{\kappa \gamma} S_{\gamma \theta}}{S_{\lambda \mu} S_{\lambda \mu}}
\propto \frac{Tr(S^3)}{Tr(S^2)}
\end{equation}
where $S_{ij} = \frac{1}{2} (\partial_i u_j + \partial_j u_i )$ is the strain-rate tensor. The above formula suggests a similarity with the Smagorinsky SGS model~\cite{Smagorinsky1963} $\nu_{\text{t}} = C_{smago} \Delta x^2 \sqrt{ S_{\theta \kappa} S_{\theta \kappa} } \propto \sqrt{ Tr(S^2)}$ while allowing back-scatter as it can change sign.\\
In order to quantify the validity of the ELBM methodology as a LES turbulence SGS model, one needs to be able to evaluate and understand the physics it implies. Firstly, one needs to control the hydrodynamic recovery and {\color{black} determine to which accuracy the Navier-Stokes equations are recovered as a function of the analyzing sub-volume size~\cite{Biferale2010}}. This is an unquestionable prerequisite. Secondly, one needs to further study the subgrid-scale model implied by the ELBM. Based on this philosophy, in this paper we propose a tool to numerically evaluate the Navier-Stokes hydrodynamic recovery of fluid flow simulations in the context of isotropic homogeneous turbulence. This tool is based on the systematic calculation of each term of the kinetic energy and enstrophy balance equations averaged over a suitable ensemble of sub-volumes of the computational grid. {\color{black}A similar approach to characterize LBM hydrodynamics was successfully used in ~\cite{Bosch,Dorschner2016} by estimating the input viscosity $\nu_0$ from the incompressible energy and enstrophy equations averaged over the whole volume. Here, we define an error with respect to an exact balance of the equation of motion and conduct a statistical analysis over sub-volumes of different sizes to assess the locality of the hydrodynamic recovery.}
The paper is organized as follows: in section~\ref{sec:2} we introduce the balance equations, their averaged counterparts over a sub-volume $V$ and we define balancing errors as a measure of the hydrodynamic recovery; in section~\ref{sec:3} we present the numerical set-up for the simulations of 2D isotropic homogeneous turbulence and for the statistical analysis of the balancing errors; in section~\ref{sec:4} we present a validation of the tool by comparing the hydrodynamic recovery of an ensemble of LBGK simulations to an ensemble of Pseudo-Spectral (PS) simulations in the case of decaying flows; in section~\ref{sec:5} we benchmark the tool on LBGK simulations of forced turbulence for a range of {\color{black} increasing Reynolds numbers}, {\color{black} while linking the results to  the corresponding statistics of the Mach number;} some concluding remarks will follow in section~\ref{sec:6}.

\section{Hydrodynamic recovery for energy and enstrophy balance in 2D}\label{sec:2}

In order to characterize the hydrodynamic recovery of a simulation, we calculate the average over sub-volumes of the terms in both the kinetic energy and the enstrophy balance equations. Starting from the formulation of the macroscopic LBM momentum conservation (see Eq.~\eqref{eq:N-S_LBM}) and mass conservation (see Eq.~\eqref{eq:mass_LBM}), one can obtain the kinetic energy ($E = \frac{\rho u_i u_i}{2}$) balance equation and the enstrophy ($\Omega = \frac{\omega_i \omega_i}{2}$, with $\omega_i$ the component of the vorticity $\vec{\omega} = \vec{\nabla} \times \vec{u}$ along $\vec{e}_i$) balance equation
\begin{align} \label{eq:E_balance}
  \begin{split}
    \partial_t \left(\frac{\rho u_i u_i}{2}\right)
    =& - u_i \partial_i p - \nu \rho \left( \partial_j u_i
    + \partial_i u_j \right) \partial_j u_i + u_i F_i \\
    & - \partial_j \left(\frac{\rho u_i u_i}{2} u_j\right)
    + \partial_j \left(\nu \rho u_i \left( \partial_j u_i + \partial_i u_j \right)\right)
  \end{split}
\end{align}
\begin{align} \label{eq:Z_balance}
  \begin{split}
    \partial_t \left(\frac{\omega_i \omega_i}{2}\right)
    =&- \partial_j \left(\frac{\omega_i \omega_i}{2} u_j\right)
    + \omega_i \omega_j \partial_j u_i
    + H_i (\nu) \epsilon_{i j k} \partial_j \omega_k
    + \omega_i \epsilon_{i j k} \partial_j \left(\frac{1}{\rho} F_k \right)\\
    & - \partial_j \left(\frac{\omega_i \omega_i}{2} u_j\right)
    + \partial_j \left(\epsilon_{i j k} \omega_i H_k (\nu)\right)
  \end{split}
\end{align}
where $\epsilon$ is the Levi-Civita symbol and $H_i (\nu)= \frac{1}{\rho} \partial_j \nu \rho$ $\left( \partial_i u_j + \partial_j u_i \right)$. Equations~\eqref{eq:E_balance} and~\eqref{eq:Z_balance} are locally valid. The next step is to calculate the average of each term of the balance equations over {\color{black} a sub-volume $V$}
\begin{align}
  \label{eq:E_balance_average}
  \begin{split}
    LHS_V^E =& \, \partial_t \big \langle \frac{\rho u_i u_i}{2} \big \rangle_V\\
    =& - \big \langle \partial_j \left(\frac{\rho u_i u_i}{2} u_j\right)  \big \rangle_V
    - \big \langle u_i \partial_i p \big \rangle_V
    + \big \langle u_i F_i \big \rangle_V\\
    &-  \big \langle \nu \rho \left( \partial_j u_i + \partial_i u_j \right) \partial_j u_i \big \rangle_V
    +  \big \langle \partial_j \left(\nu \rho u_i \left( \partial_j u_i + \partial_i u_j \right)\right) \big \rangle_V
    \\
    =& \, RHS^{E, \, 1}_V + RHS^{E, \, 2}_V + RHS^{E, \, 3}_V+ RHS^{E, \, 4}_V + RHS^{E, \, 5}_V\\
    =& \, RHS_V^E
  \end{split}
\end{align}
\begin{align}
  \label{eq:Z_balance_average}
  \begin{split}
    LHS_V^{\Omega} =& \, \partial_t \big \langle \frac{\omega_i \omega_i}{2} \big \rangle_V\\
    =& - \big \langle \partial_j \left(\frac{\omega_i \omega_i}{2} u_j\right) \big \rangle_V
    - \big \langle \frac{\omega_i \omega_i}{2} \partial_j u_j \big \rangle_V
    + \big \langle \omega_i \epsilon_{i j k} \partial_j \left(\frac{1}{\rho} F_k \right) \big \rangle_V\\
    &+ \big \langle H_i (\nu) \epsilon_{i j k} \partial_j \omega_k \big \rangle_V
    + \big \langle \partial_j \left(\epsilon_{i j k} \omega_i H_k(\nu)\right) \big \rangle_V
    + \big \langle \omega_i \omega_j \partial_j u_i \big \rangle_V\\
    =& \, RHS^{\Omega, \, 1}_V + RHS^{\Omega, \, 2}_V + RHS^{\Omega, \, 3}_V + RHS^{\Omega, \, 4}_V
    + RHS^{\Omega, \, 5}_V + RHS^{\Omega, \, 6}_V\\
    =& \, RHS^{\Omega}_V
  \end{split}
\end{align}
where $ \big \langle \cdots \big \rangle_V$ denotes the average over a generic volume $V$. Equations~\eqref{eq:E_balance_average} and~\eqref{eq:Z_balance_average} describe the physical balance between the time derivative of the averaged energy and enstrophy ($LHS^{E, \, \Omega}_V$) and the right-hand side ($RHS^{E, \, \Omega}_V$) comprising all the physical contributions responsible for their evolution: the effect of compressibility, dissipation, input, and the transport and diffusive fluxes. It is worth pointing out that equations \eqref{eq:E_balance_average} and \eqref{eq:Z_balance_average} remain valid for a viscosity changing in space and time $\nu = \nu_{\text{eff}} (\vec{x}, \, t)= \nu_0 + \nu_{\text{t}} (\vec{x}, \, t)$. Notice that in 3D, the enstrophy balance must include another additional term stemming from vortex stretching~\cite{Davidson2015}.\\
To measure the accuracy of the hydrodynamic recovery over a sub-volume $V$, we define a balancing error for the kinetic energy and enstrophy balance, $\delta_V^{E}$ and $\delta_V^{\Omega}$ respectively. At a time $t$, $\delta^{E, \, \Omega}_V(t)$ is obtained by dividing the absolute difference between the $RHS^{E, \, \Omega}_V(t)$ and the $LHS^{E, \, \Omega}_V(t)$ terms by the term of the right-hand side with the maximum absolute value {\it i.e.}
\begin{equation}\label{eq:delta_E}
\delta_{V}^{E}(t)=\frac{\left| RHS_{V}^{E}(t)-LHS_{V}^{E}(t) \right|}{\max_i \left| RHS_{V}^{E, \, i}(t) \right|}
\end{equation}
and
\begin{equation}\label{eq:delta_Z}
\delta_{V}^{\Omega}(t)=\frac{\left| RHS_{V}^{\Omega}(t)-LHS_{V}^{\Omega}(t) \right|}{\max_i \left| RHS_{V}^{\Omega, \, i}(t) \right|}.
\end{equation}
If for a sub-volume $V$ at a time $t$ the balance equations are perfectly respected on average, we must have $\delta_{V}^{E}(t) \equiv \delta_{V}^{\Omega}(t) \equiv 0$.
\section{Numerical set-up for the statistical analysis of 2D homogeneous isotropic turbulence hydrodynamics}\label{sec:3}

\begin{figure}[H]
\centering
\includegraphics[width=14cm]{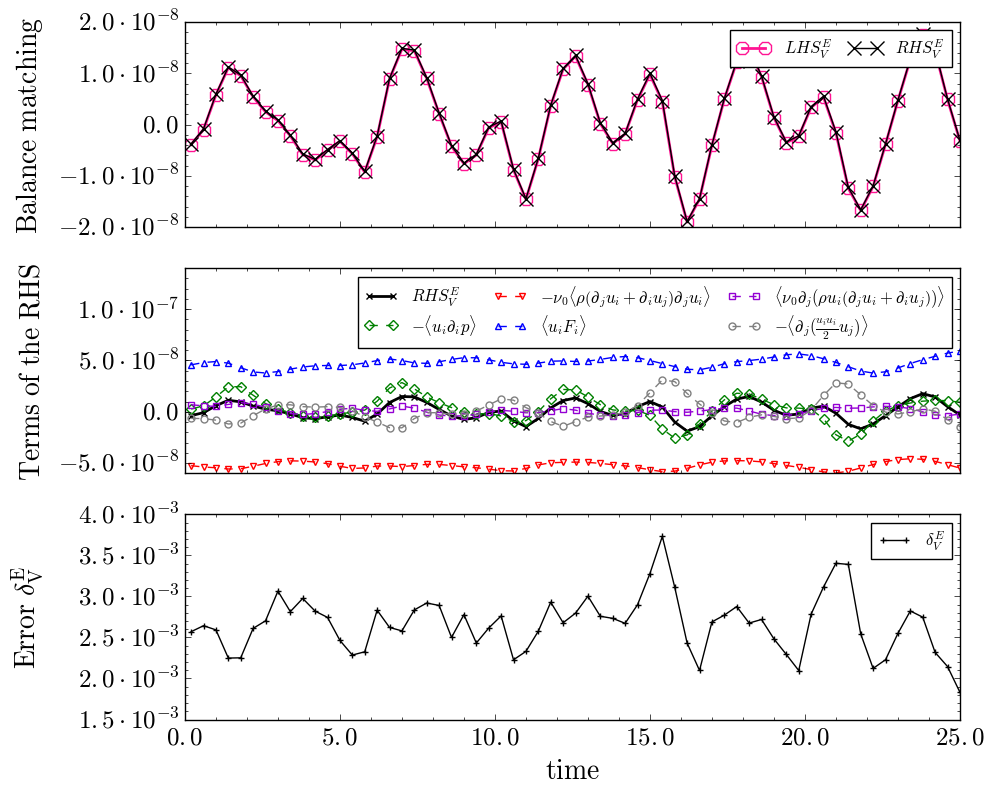}
\caption{Typical time-evolution of the kinetic energy balancing over a single sub-volume of size {\color{black} $181 \times 181$ shown for a forced LBGK simulation with $\tau_0 = 0.60$ ($Re \approx 90$)} on a $256 \times 256$ grid. The top figure shows the matching between the $LHS^{E}_V$ and the $RHS^{E}_V$, the middle figure shows the contribution of each $RHS^{E, \, i}_V$ term and their sum $RHS^{E}_V$, and the bottom figure shows the balancing error $\delta^{E}_V$.}\label{fig:example_E_balance}
\end{figure}

\begin{figure}[H]
\centering
\includegraphics[width=14cm]{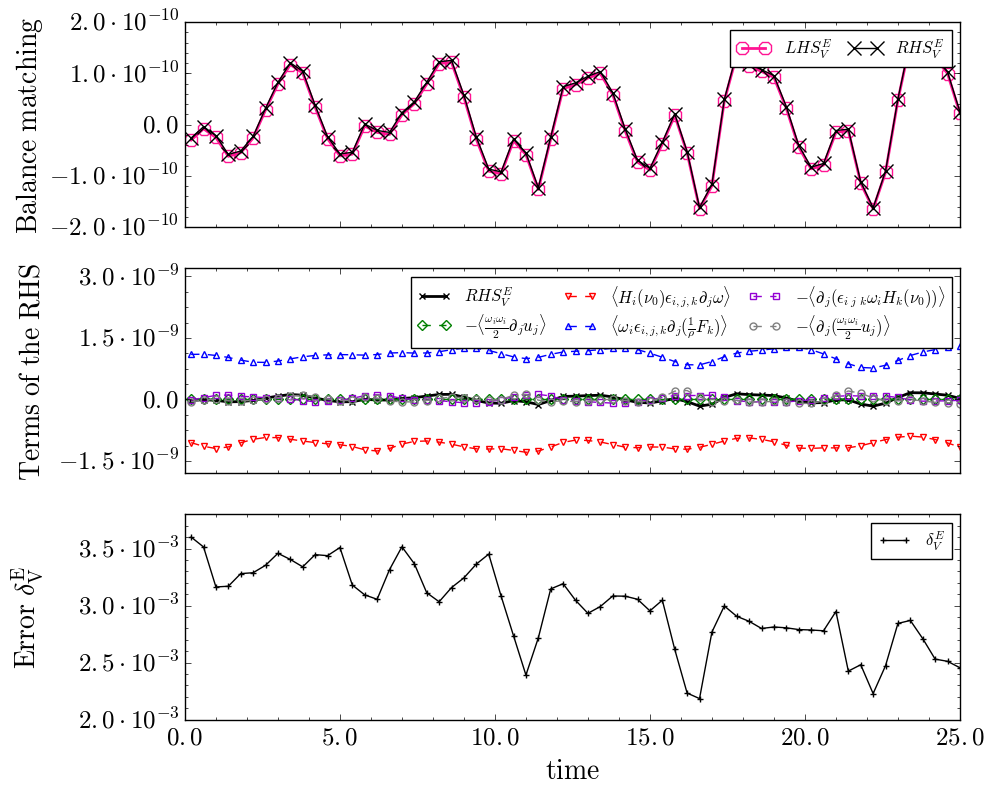}
\caption{Typical time-evolution of the enstrophy balancing over a single sub-volume of size {\color{black} $181 \times 181$ shown for a forced LBGK simulation with $\tau_0 = 0.60$ ($Re \approx 90$)} on a $256 \times 256$ grid. The top figure shows the matching between the $LHS^{\Omega}_V$ and the $RHS^{\Omega}_V$, the middle figure shows the contribution of each $RHS^{\Omega, \, i}_V$ term and their sum $RHS^{\Omega}_V$, and the bottom figure shows the balancing error $\delta^{\Omega}_V$.}\label{fig:example_Z_balance}
\end{figure}

\begin{figure}[H]
\noindent \begin{centering}
\includegraphics[width=7cm]{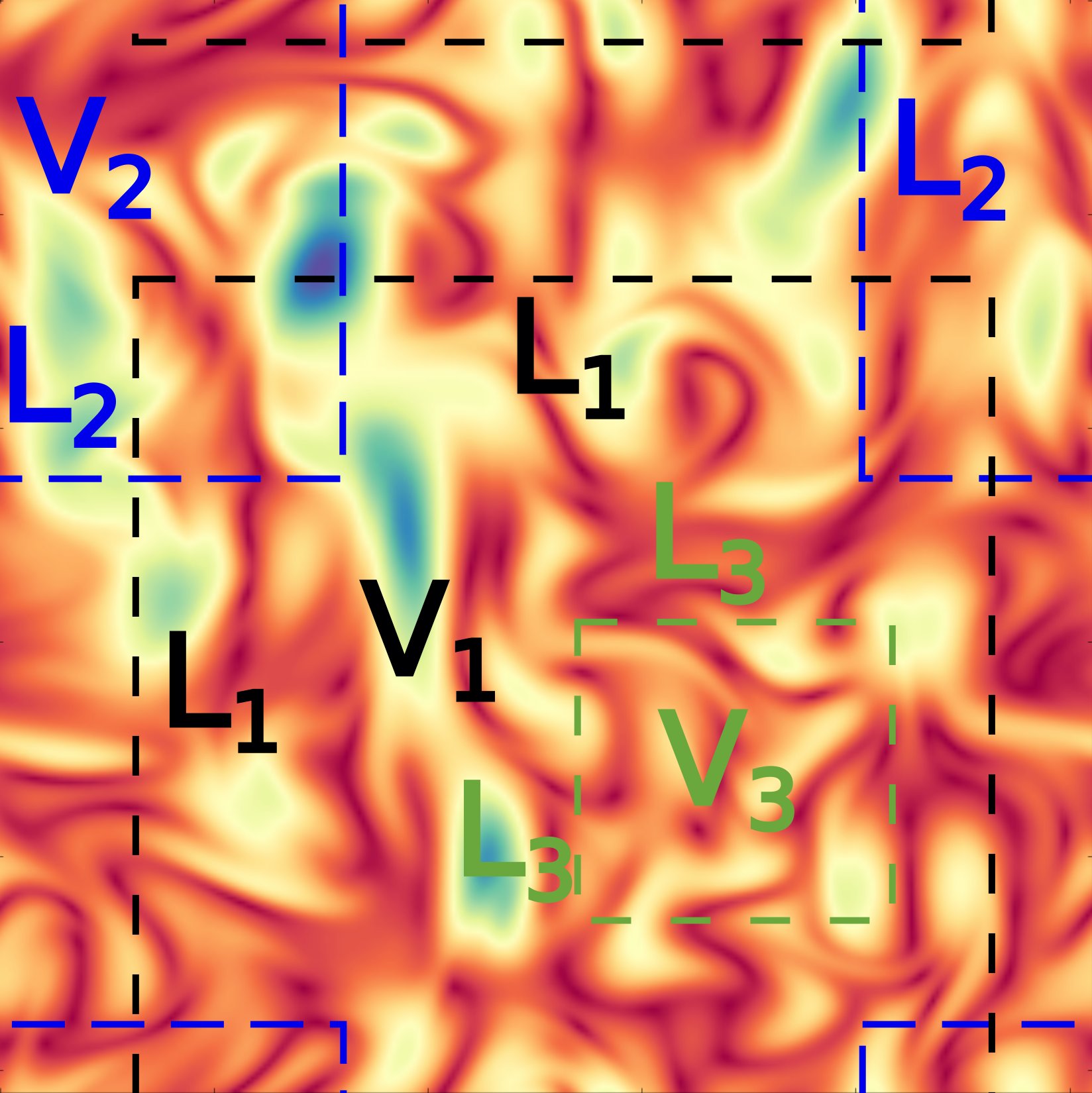}
\par\end{centering}
\caption{Illustration on a snapshot of the vorticity field of three random squared sub-volumes $V_1=L_1 \times L_1$, $V_2=L_2 \times L_2$, and $V_3=L_3 \times L_3$ {\color{black} corresponding to the sub-volume size $L_1$, $L_2$, and $L_3$ respectively.}}
\label{fig:sub-volume_example}
\end{figure}

To validate this hydrodynamic recovery check tool, we apply it to configurations obtained from simulations conducted on a periodic two-dimensional $256 \times 256$ computational grid. Turbulence is triggered by a homogeneous isotropic forcing with a constant phase $\phi$ on a shell of (dimensionless) wavenumbers $\vec{k}$ of magnitude from 5 to 7 given in a stream-function formulation
\begin{equation}\label{eq:forcing_turbulence_streamfunction}
F^{T}_{\Psi} (\vec{x})=F^{T}_{0}\sum_{5 \leq \|\vec{k}\| \leq 7}\cos\left(\frac{2\,\pi}{256}\vec{k} \cdot \vec{x}+\phi\right).
\end{equation}
The corresponding force is then obtained by taking
\begin{equation}\label{eq:forcing_turbulence}
F^{T}_x = \partial_y F^{T}_{\Psi} \qquad \text{and} \qquad F^{T}_y = -\partial_x F^{T}_{\Psi},
\end{equation}
which ensures that it does not input any incompressibility in the system as $\vec{\nabla} \cdot \vec{F}^{T} \equiv 0$. We use this forcing to define a time scale  $T_f = \sqrt{ \frac{2 \pi}{k_f F^T_{0}} }$, where $k_f$ is taken equal to six. To have some control on the Mach number and limit the effect of the backward energy cascade, characteristic of 2D turbulence~\cite{Boffetta2012, Frisch1995}, we introduce a spectral forcing to damp large-scale energy
\begin{equation}\label{eq:forcing_damping}
\vec{F}^{R} \left( \vec{x}, \, t \right) = - F^{R}_{0} \sum_{1 \leq \|\vec{k}\| \leq 2} \vec{\hat{u}} (\vec{k}, \, t) \, e^{\frac{2\,\pi}{256}\vec{k} \cdot \vec{x} }
\end{equation}
where $\vec{\hat{u}} (\vec{k}, \, t)$ is the Fourier transform of $\vec{u} (\vec{x}, \, t)$. The forcing amplitudes are fixed for all simulations to $F^{T}_{0} = 0.0008$ and $F^{R}_{0}=0.00001$. LBGK simulations are conducted on a 2D lattice with 9 discrete velocities, the D2Q9~\cite{Succi2001, Wolf-Gladrow2000, Kruger2017}, on which forcings are implemented using the exact-difference method forcing scheme~\cite{Kuperstokh2004}. The sub-volume averaged terms are calculated offline based on the {\color{black} output} configuration fields. {\color{black} A 2\textsuperscript{nd} order explicit Euler scheme is used to evaluate time derivatives, while a 8\textsuperscript{th} order centered scheme is applied for the space-derivatives, respectively
\begin{equation}
\left. \frac{\partial \mathbf{A}}{\partial t}\right |^n_{i,j}
\sim
\frac{3 A_{i,j}^{n}-4 A_{i,j}^{n-1}+A_{i,j}^{n-2}}{2 \,\Delta t}
\label{eq:discretisation_t2} \text{, and }
\end{equation}

\begin{align}
\begin{split}
\left. \frac{\partial \mathbf{A}}{\partial x} \right|^n_{i,j}
\sim &
\frac{- \frac{1}{56} A^n_{i+4,j} + \frac{4}{21} A^n_{i+3,j} - A^n_{i+2,j} + 4 A^n_{i+1,j}
- 4 A^n_{i-1,j} + A^n_{i-2,j} - \frac{4}{21} A^n_{i-3,j} + \frac{1}{56} A^n_{i-4,j}} {5  \,\Delta x}\\
\text{\& } \left. \frac{\partial \mathbf{A}}{\partial y} \right|^n_{i,j}
\sim &
\frac{- \frac{1}{56} A^n_{i,j+4} + \frac{4}{21} A^n_{i,j+3} - A^n_{i,j+2} + 4 A^n_{i,j+1}
- 4 A^n_{i,j-1} + A^n_{i,j-2} - \frac{4}{21} A^n_{i,j-3} + \frac{1}{56} A^n_{i,j-4}} {5  \,\Delta y} \text{.}
\end{split}
\label{eq:dscretisation_s8}
\end{align}
}
Examples of the balancing of the terms of the energy and enstrophy equations are illustrated in Figs.~\ref{fig:example_E_balance} and~\ref{fig:example_Z_balance} respectively. {\color{black} In both cases, the matching between the left-hand side ($LHS^{E, \, \Omega}_V$) and the right-hand side ($RHS^{E, \, \Omega}_V$) highlights very small discrepancies observed. Typically, the total $RHS^{E, \, \Omega}_V$ terms are the result of the sum of significantly higher amplitude terms. Eventually, the resulting balancing errors $\delta^{E, \, \Omega}_V$ is of the order ${\cal O} (10^{-3})$ for both the kinetic energy balancing and the enstrophy balancing, resulting in an excellent hydrodynamic recovery.}
{\color{black}In order to gather statistics of both balancing errors $\delta_{V}^{E, \, \Omega}(t)$  for a given sub-volume size $L$, we calculate them over squared sub-volumes $V= L \times L$ randomly chosen in space as illustrated in Fig.~\ref{fig:sub-volume_example}}.
{\color{black} To present the results, we introduce the normalized sub-volume size $l=\frac{L}{L_0}$ with $L_0 = 256$ the size of the squared computational domain, and we group together the balancing errors $\delta^{E, \, \Omega}_{l}(t) = \delta^{E, \, \Omega}_{V = L \times L}(t)$ obtained for all sub-volumes of the same normalized sub-volume size $l$ on the same configuration at time $t$. We conduct a statistical analysis and define their mean $\mu^{E, \, \Omega}_{l}(t)$ and their standard deviation $\sigma^{E, \, \Omega}_{l}(t)$. The number of sub-volumes processed for a normalized sub-volume size $l$ is shown in Table~\ref{tab:number_subvolumes}}.
\begin{table}[H]
\centering
\begin{tabular}{ @{} c c c @{} }
\toprule[1pt]
Sub-volume size $L$ & {\color{black}Corresponding normalized sub-volume size $l$} & Number of sub-volumes processed \\
\midrule[0.5pt]
\addlinespace[0.5mm]
$L = 256$     & {\color{black}$l=1$} & 1  \\
$100 \leq L < 256$ & {\color{black}$0.4 \leq l < 1$} & 1000  \\
$10 \leq L < 100$ & {\color{black}$0.04 \leq l < 0.4$} & 5000  \\
$L < 10$ & {\color{black}$l < 0.04$} & 10000 \\
\bottomrule[1pt]
\end{tabular}
\caption{{\color{black} Number of sub-volumes processed per sub-volume size $L$}}
\label{tab:number_subvolumes}
\end{table}

\section{Validation: LBGK against Pseudo-Spectral on an ensemble of decaying flow simulations}\label{sec:4}

\begin{figure}[H]
\centering
\includegraphics[height=5cm]{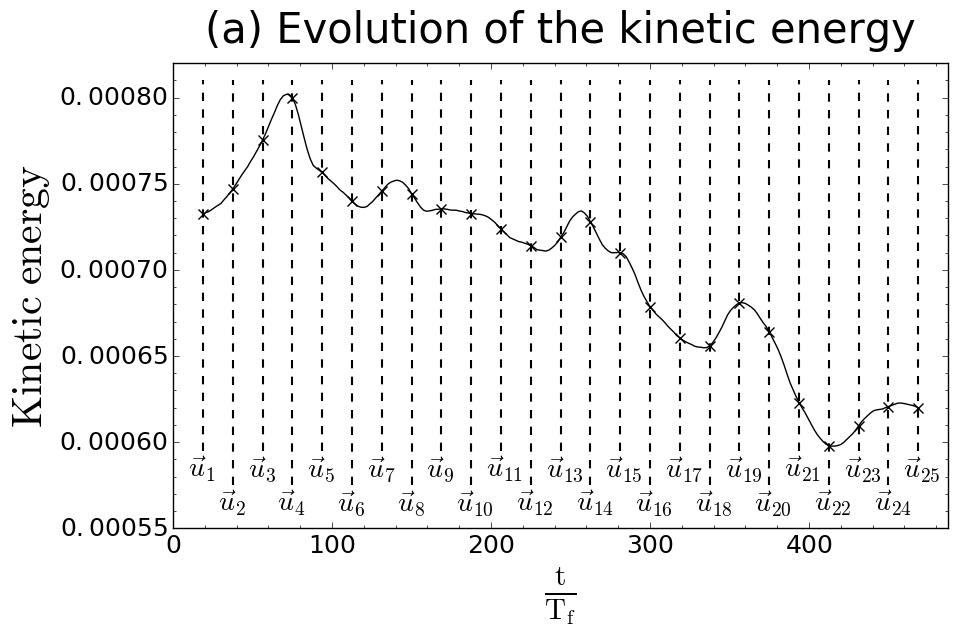}
\includegraphics[height=5cm]{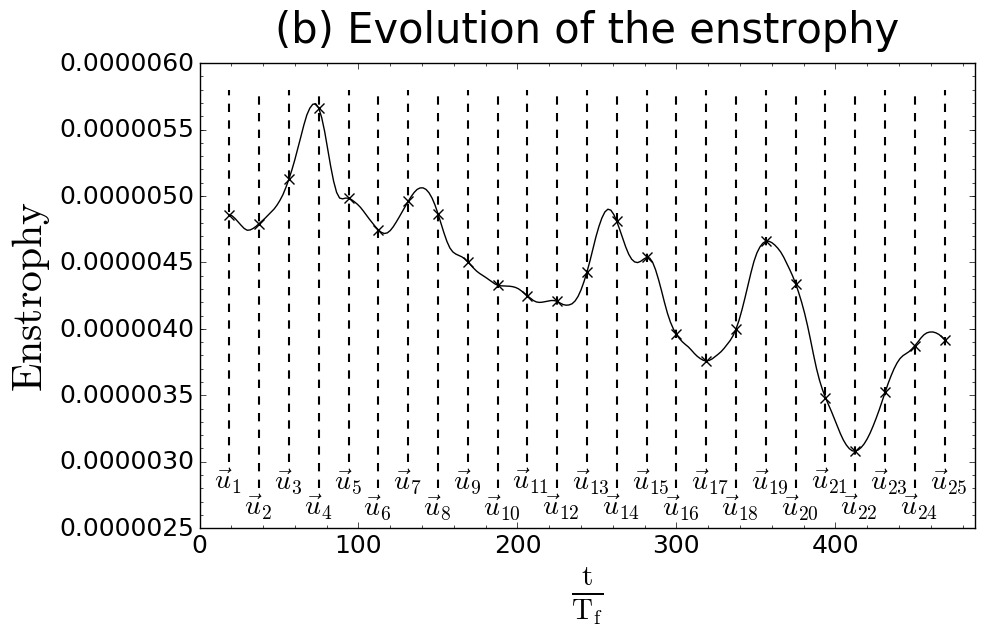}
\caption{Evolution of the kinetic energy (a) and of the enstrophy (b) of the forced LBGK simulation. The 25 vertical lines highlight the sampled configurations used to initialize the 25 decaying flow simulations of the PS and the LBGK ensembles.}
\label{fig:sampling_PS}
\end{figure}

\begin{figure}[H]
\centering
\includegraphics[width=10cm]{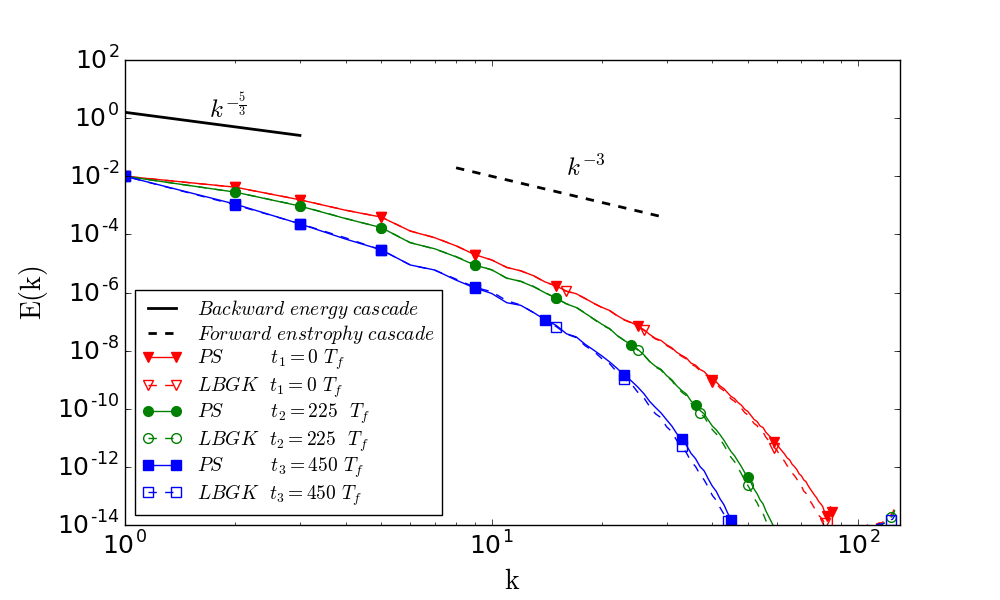}
\caption{Superposed ensemble-averaged energy spectrum shown for three selected time instances for the PS and the LBGK simulations.}
\label{fig:spectra_PS}
\end{figure}

\begin{figure}[H]
\centering
\includegraphics[width=0.49\textwidth]{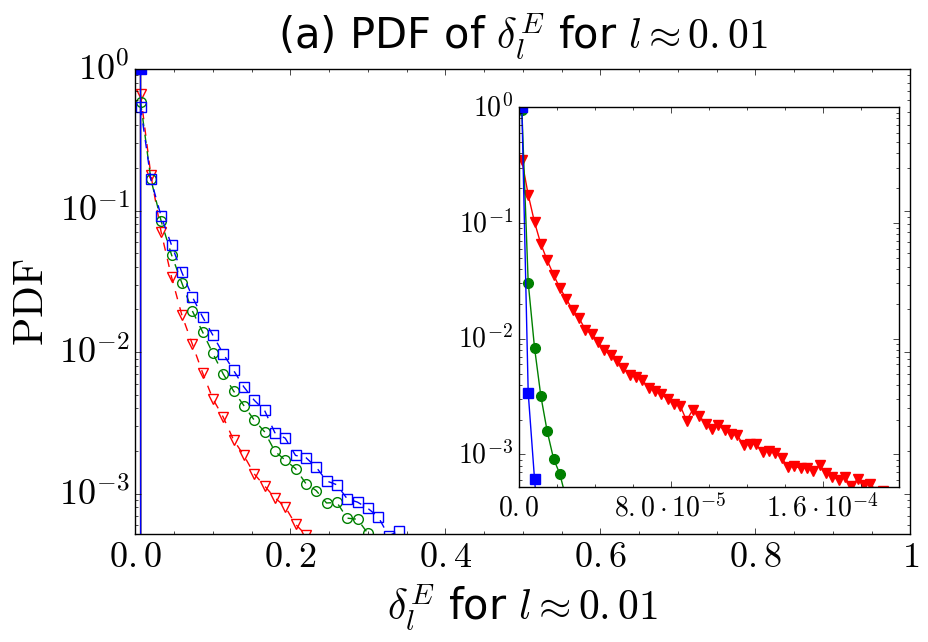}
\hfill
\includegraphics[width=0.49\textwidth]{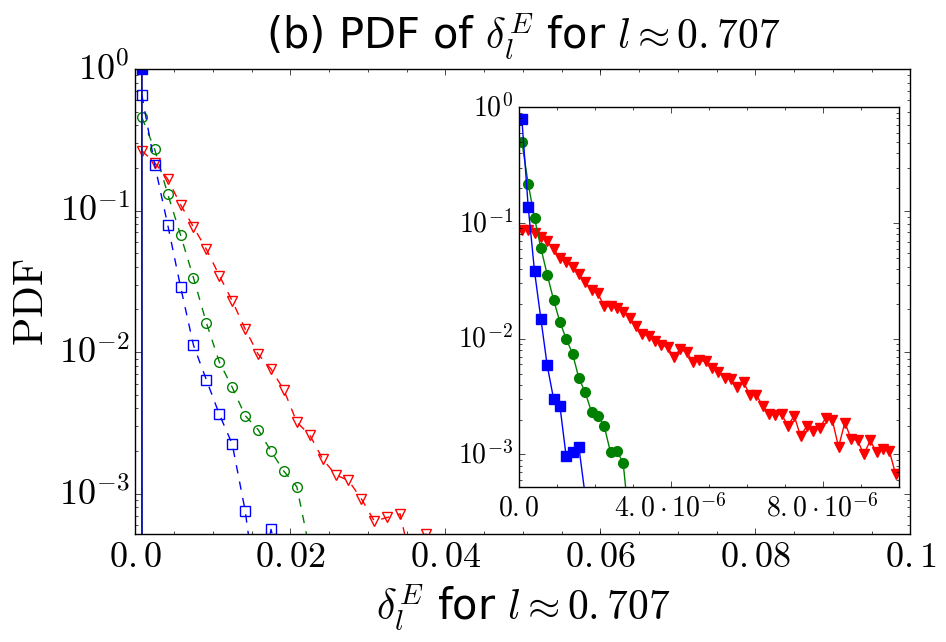}
\vspace*{0.3cm}
\includegraphics[width=0.49\textwidth]{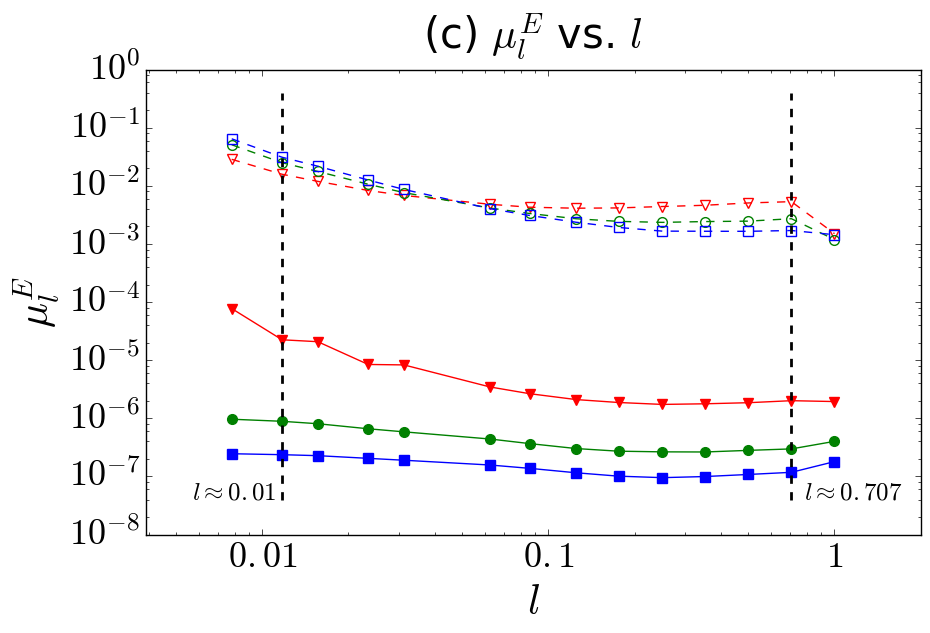}
\hfill
\includegraphics[width=0.49\textwidth]{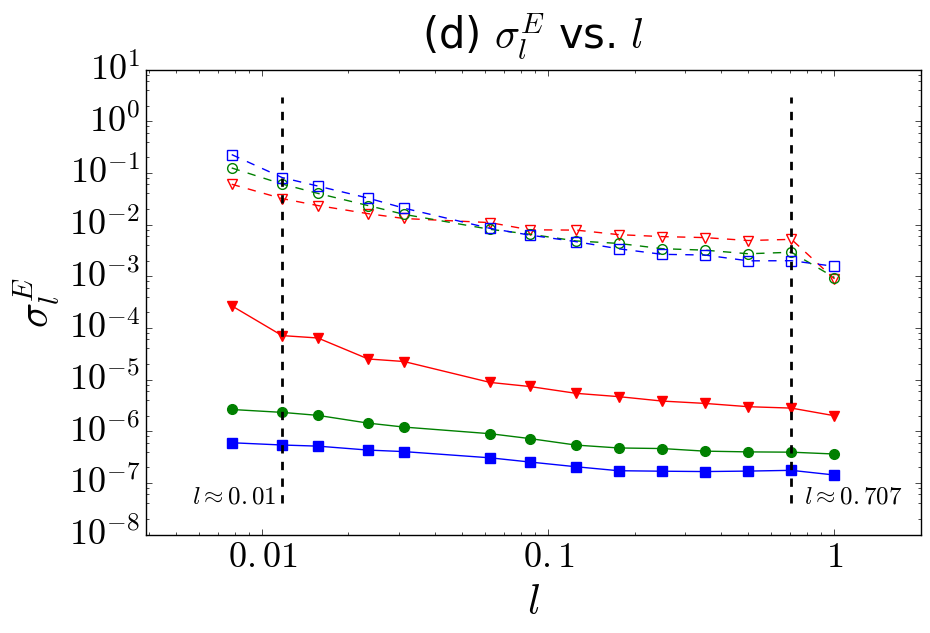}

\includegraphics[width=0.9\textwidth]{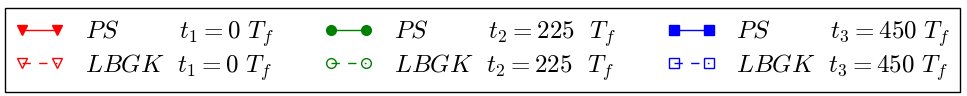}
\caption{Statistics of the balancing error obtained from the kinetic energy balance {\color{black}$\delta_l^{E}$ (see Eq.~\eqref{eq:delta_E}) against the normalized size of the sub-volume $l$ shown for the PS and LBGK ensemble of 25 decaying simulations for three selected times. Top figures are PDF of the balancing error for sub-volumes corresponding to $l \approx 0.01$ (Panel (a)) and $l \approx 0.707$ (Panel (b)) and insets shows the PDFs of the balancing error for the PS ensemble alone}. Bottom figures are the mean (Panel (c)) and the standard deviation (Panel (d)) of the balancing error.\label{fig:results_E_PS-LBGK}}
\end{figure}

\begin{figure}[H]
\centering
\includegraphics[width=0.49\textwidth]{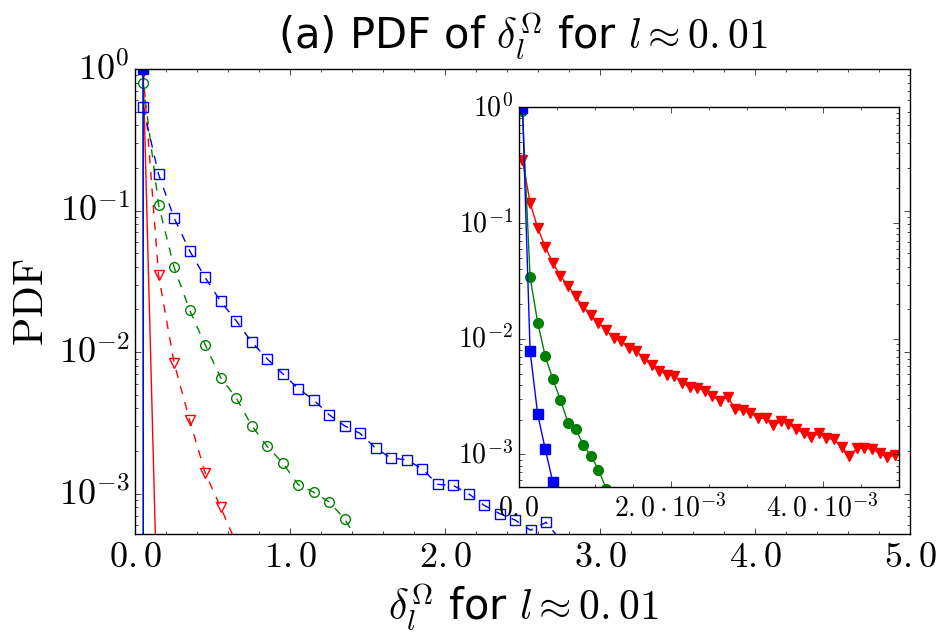}
\hfill
\includegraphics[width=0.49\textwidth]{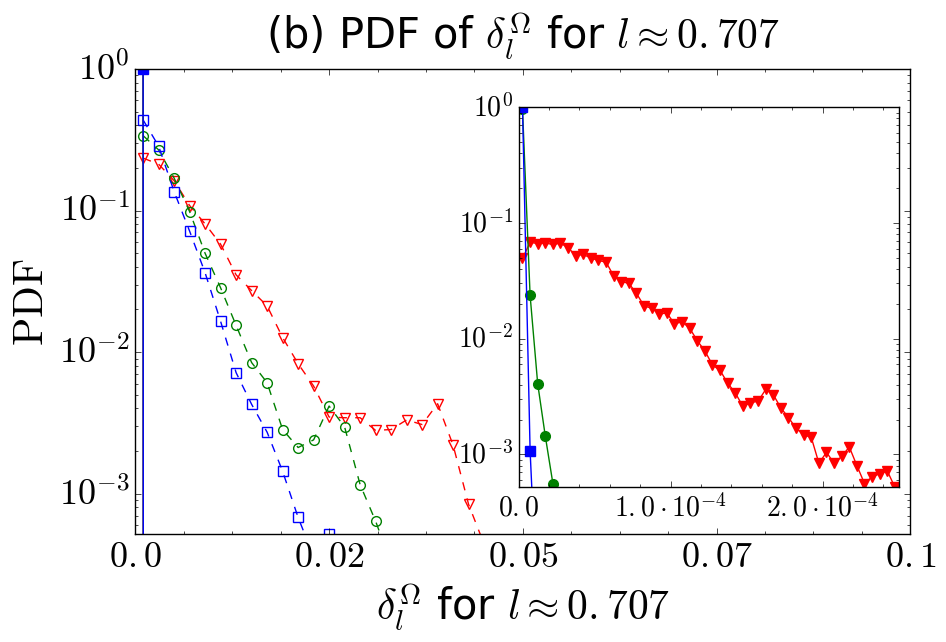}
\vspace*{0.3cm}
\includegraphics[width=0.49\textwidth]{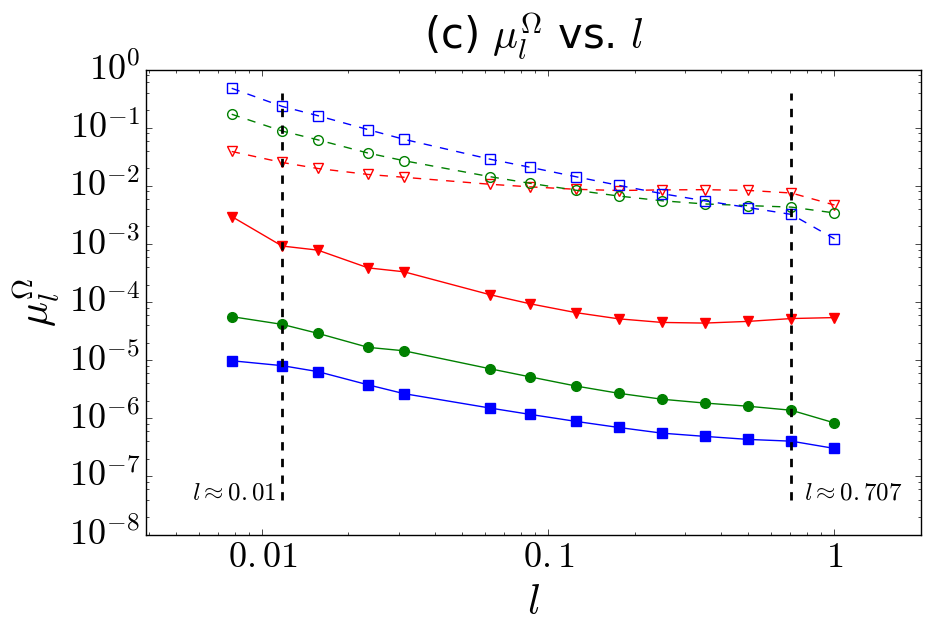}
\hfill
\includegraphics[width=0.49\textwidth]{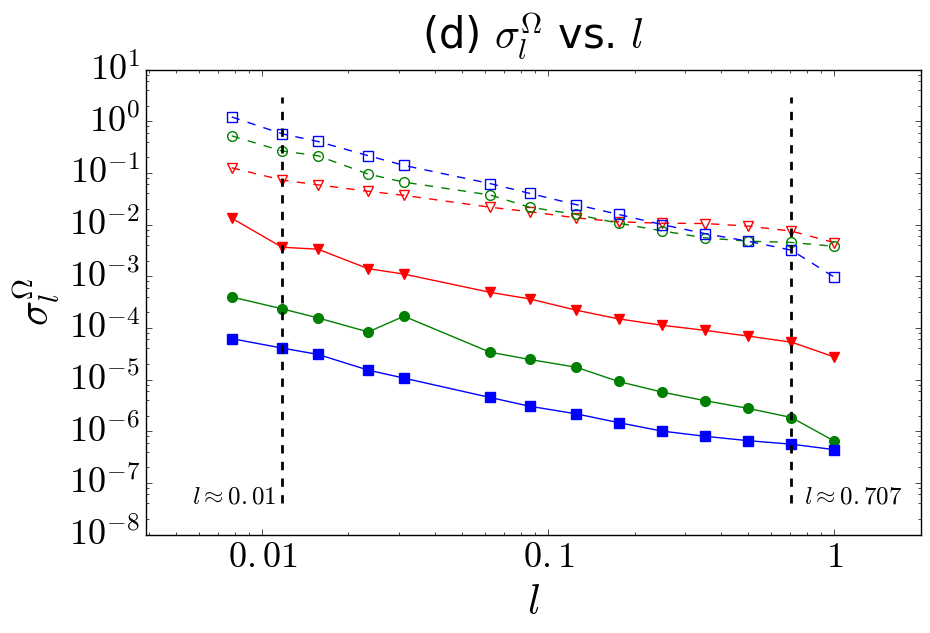}

\includegraphics[width=0.9\textwidth]{legend_PS-LBGK}
\caption{Statistics of the balancing error obtained from the enstrophy balance {\color{black}$\delta_l^{\Omega}$ (see Eq.~\eqref{eq:delta_Z}) against the normalized size of the sub-volume $l$ shown for the PS and LBGK ensemble of 25 decaying simulations for three selected times. Top figures are PDFs of the balancing error for sub-volumes corresponding to $l \approx 0.01$ (Panel (a)) and $l \approx 0.707$ (Panel (b)) and insets shows the PDFs of the balancing error for the PS ensemble alone}. Bottom figures are the mean (Panel (c)) and the standard deviation (Panel (d)) of the balancing error.\label{fig:results_Z_PS-LBGK}}
\end{figure}

\begin{figure}[H]
\centering
\includegraphics[width=0.49\textwidth]{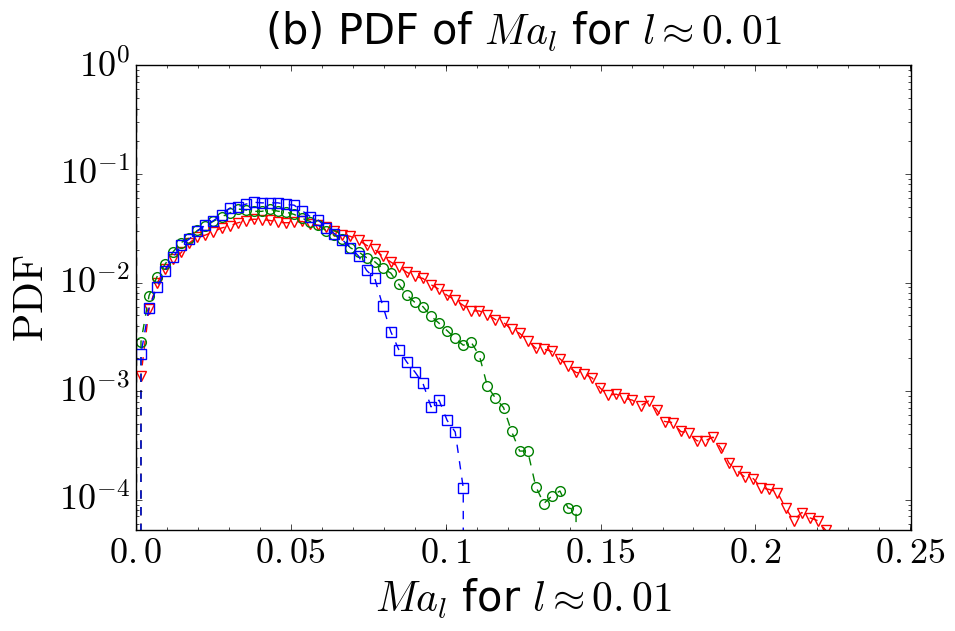}
\hfill
\includegraphics[width=0.49\textwidth]{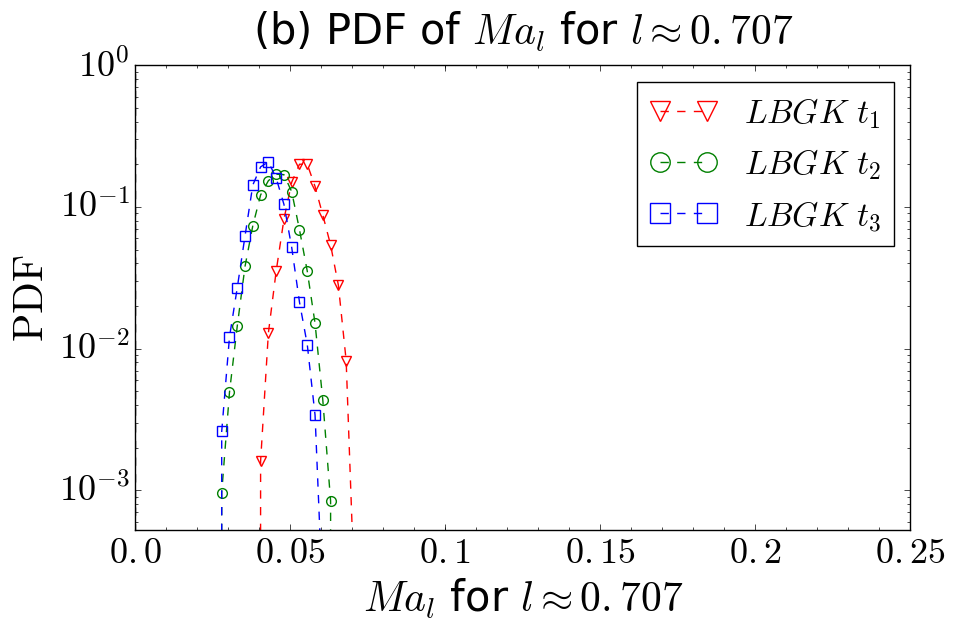}
\vspace*{0.3cm}
\includegraphics[width=0.49\textwidth]{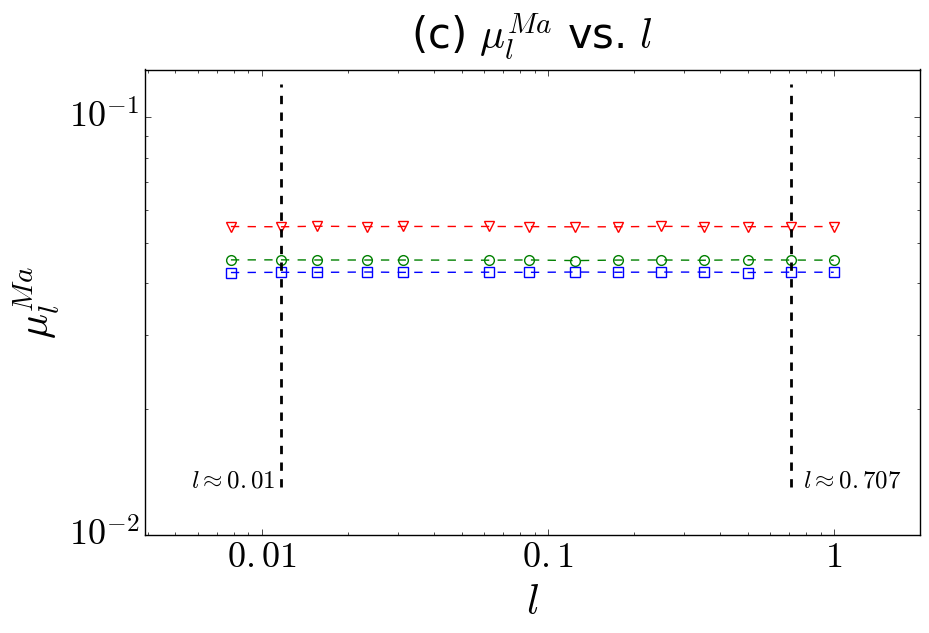}
\hfill
\includegraphics[width=0.49\textwidth]{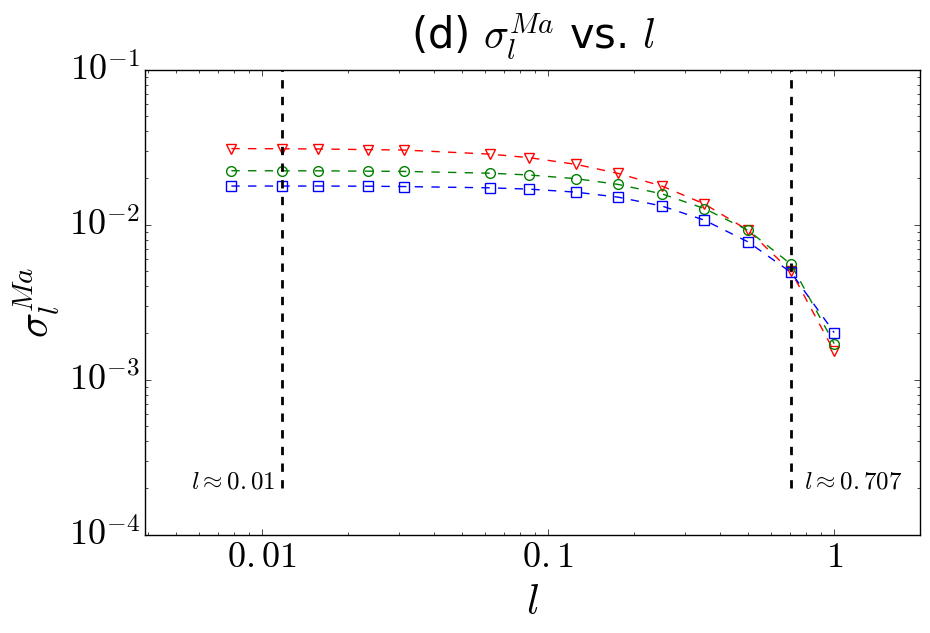}
\caption{Statistics of the Mach number at {\color{black} normalized sub-volume size $l$ (see Eq.~\eqref{eq:MaL}) $Ma_l$ against the normalized size of the sub-volume $l$ shown for the LBGK ensemble of 25 decaying simulations for three selected times. Top figures are PDFs of $Ma_l$ for sub-volumes corresponding to $l\approx 0.01$ (Panel (a)) and $l \approx 0.707$ (Panel (b)). Bottom figures are the mean (Panel (c)) and the standard deviation (Panel (d)) of $Ma_l$}.\label{fig:mach_decaying_LBGK}}
\end{figure}

To understand how LBGK recovers hydrodynamics, we compare the statistics of the balancing errors obtained from LBGK simulations to the one obtained from PS simulations, which are used as a reference. To this aim we generate ensembles of LBGK and PS simulations: we conduct a statistically {\color{black} stationary forced LBGK $Re \approx 1200$ ($\tau_0 = 0.52$)} simulation that we sample into 25 configurations as shown in Fig.~\ref{fig:sampling_PS}, the number 25 being chosen in order to recover smooth statistics. {\color{black} Each of those configurations is then used to restart a LBGK simulation and to compute the corresponding vector potential $\vec{b}$ such as $\vec{u} = \vec{\nabla} \times \vec{b}$ to initialize an incompressible PS simulation} at the same Reynolds number, thus ensuring that they solve the same physics. Specifically, we set
\begin{equation}\label{eq:reynolds_PS-LBGK}
Re=\frac{U^{LBGK}_{RMS} L^{LBGK}}{\nu^{LBGK}_0}=\frac{U^{PS}_{RMS} L^{PS}}{\nu^{PS}_0}
\end{equation}
with $U^{PS}_{RMS} = U^{LBGK}_{RMS}\frac{\Delta x^{LBGK}}{\Delta t^{LBGK}}$, $L^{PS} = 2 \pi = L^{LBGK} \Delta x^{LBGK}$, and $\nu_0^{PS} = \nu_0^{LBGK} \frac{(\Delta x^{LBGK})^2}{\Delta t^{LBGK}}$ and where $\nu_0^{LBGK} = c_s^2 (\tau_0 - 0.5)$ with $\tau_0 = 0.52$ in all simulations. Having fixed $\Delta x^{LBGK} = \frac{2 \pi}{256}$, $\tau_0 = 0.52$, and $\Delta t^{LBGK} = 0.001$, we obtain $\nu_0^{PS} \approx 0.004 $. We set $\Delta t^{PS} = 0.0005$ in order to be able to dump configurations of PS and LBGK simulations at the same physical time ($\Delta t^{LBGK} \propto \Delta t^{PS}$), while ensuring the stability of the PS simulations. Moreover, the velocity fields generated by the forced LBGK simulation have to be normalized by a factor $\frac{\Delta x^{LBGK}}{\Delta t^{LBGK}}$ before they are used to initialize the PS simulations. After initialization, the simulations are then left with no forcing to decay for a duration of $450 \, T_f$, where $T_f$ is the time scale based on the forcing as discussed in section \ref{sec:3}. Eventually, the superposed ensemble-averaged energy spectrum for both ensemble at three selected times $t_1= 0$, $t_2 = 225 T_f$, and $t_3 = 450 T_f$ are in very good agreement (Fig.~\ref{fig:spectra_PS}). The pressure field for the PS simulations is obtained by solving, for each configuration, the Poisson equation for pressure, while the pressure field for the LBGK simulations is obtained directly from the density field $p = c_s^2 \rho$.\\
We show the results of the statistical analysis of the kinetic energy balancing error {\color{black} $\delta_l^{E}$} and enstrophy balancing error {\color{black} $\delta_l^{\Omega}$} in Figs.~\ref{fig:results_E_PS-LBGK} and~\ref{fig:results_Z_PS-LBGK} respectively. As expected, the PS method recovers hydrodynamics with a significant higher accuracy than the LBGK, with a clear improvement with time as the Reynolds number decreases and the simulations become increasingly resolved. {\color{black} This improvement with time cannot be well appreciated in the LBGK simulations, as it appears to be sub-leading in both the energy balance statistics  $\mu_l^{E}$ and $\sigma_l^{E}$ (Fig.~\ref{fig:results_E_PS-LBGK}, Panels (c)-(d)) and the the enstrophy balance statistics  $\mu_l^{\Omega}$ and $\sigma_l^{\Omega}$ (Fig.~\ref{fig:results_Z_PS-LBGK}, Panels (c)-(d)).} Taken all together, the statistical analysis of the balancing errors {\color{black}$\delta_l^{E}$ and $\delta_l^{\Omega}$ show that hydrodynamic recovery is excellent on large sub-volumes and two orders of magnitude larger on small sub-volumes (see Figs.~\ref{fig:results_E_PS-LBGK} and~\ref{fig:results_Z_PS-LBGK}, Panels (a)-(b)), the errors remaining however of order ${\cal O} (10^{-1})$.}\\
To understand if the range of Mach numbers simulated affects the hydrodynamic recovery, we plot the statistics on the Mach number at {\color{black} the normalized sub-volume size $l$, {\it i.e.}
\begin{equation}\label{eq:MaL}
Ma_l = \big \langle \frac{U_{RMS}}{c_s} \big \rangle_{V=L \times L} \text{, } l = \frac{L}{L_0}
\end{equation}}
as shown in Fig.~\ref{fig:mach_decaying_LBGK}. We observe a steady mean (Fig.~\ref{fig:mach_decaying_LBGK}-(c)) going from about 0.55 to 0.4, and a steady standard deviation (Fig.~\ref{fig:mach_decaying_LBGK}-(d)) up to $L \approx 20$. As expected for decaying flows, the Mach number gradually decreases in time {\color{black} for all sub-volume sizes}. {\color{black} The statistical analysis of the decaying LBGK simulations is quite helpful to further assess the importance of the terms proportional to $Ma^3$ neglected in the momentum equation (see Eq.~\eqref{eq:N-S_LBM}). Indeed, if we look at the statistics of the energy and enstrophy balancing errors in Figs.~\ref{fig:results_E_PS-LBGK} and ~\ref{fig:results_Z_PS-LBGK}, we notice that if the Mach number was impacting the balancing errors, we would have observed a statistics that varies in time as the Mach number decays.} Thus, we can conclude that for the range of simulated Mach numbers the LBGK is a trustworthy Navier-Stokes solver, \textit{i.e.} the Mach number is low enough so that all higher order Mach number terms that were neglected in the momentum equation do not affect the hydrodynamics.
\section{Forced LBGK hydrodynamics}\label{sec:5}
\begin{figure}[H]
\centering
\includegraphics[height=5cm]{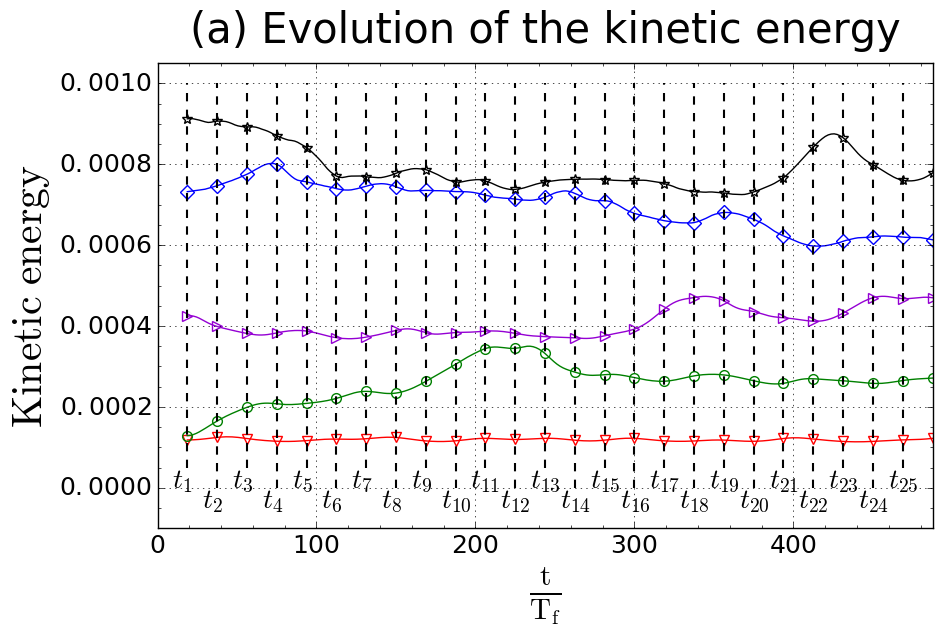}
\includegraphics[height=5cm]{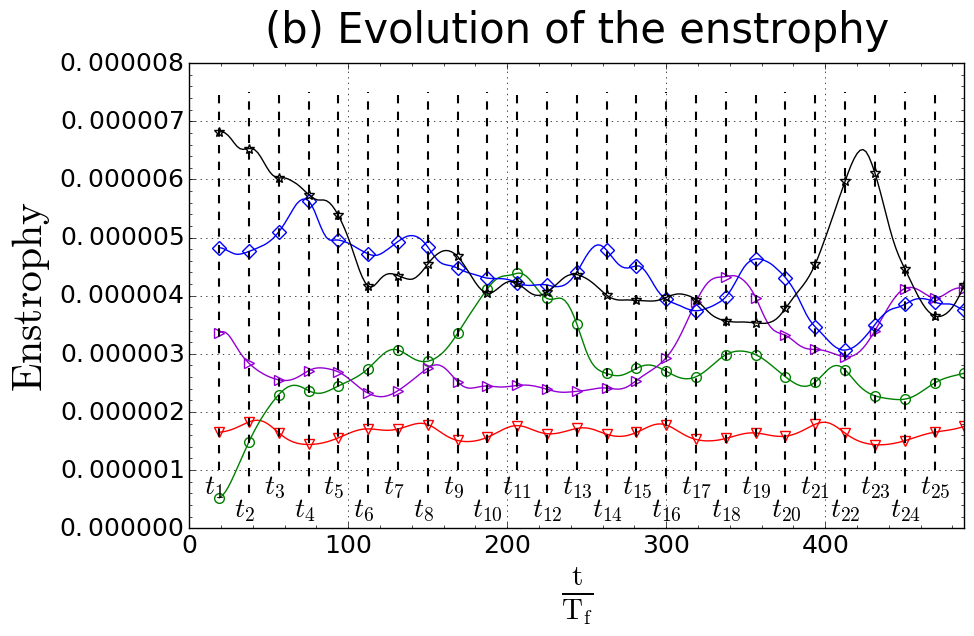}
\includegraphics[width=0.9\textwidth]{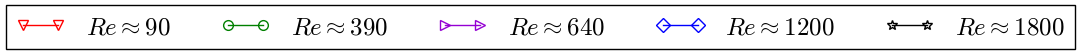}
\caption{Evolution of the kinetic energy (a) and of the enstrophy (b) of LBGK simulations for five different relaxation times. The 25 vertical lines highlight the time when configurations were processed to gather statistics in space and time of the balancing errors.}
\label{fig:sampling_LBGK}
\end{figure}

\begin{figure}[H]
\centering
\includegraphics[width=10cm]{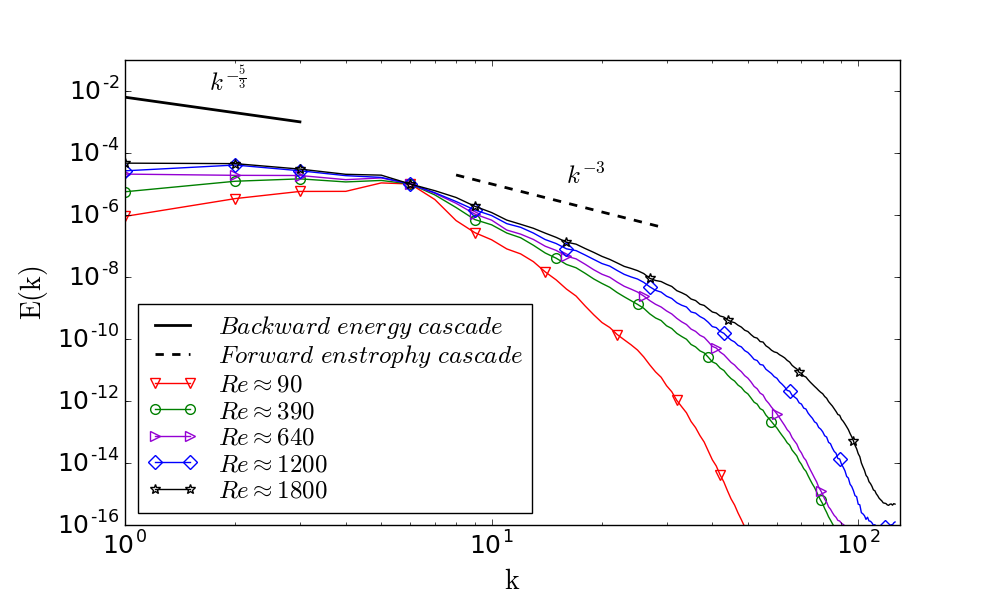}
\caption{Superposed time-averaged spectrum of LBGK simulations for five different relaxation times.}
\label{fig:spectra_LBGK}
\end{figure}

\begin{figure}[H]
\centering
\includegraphics[width=0.49\textwidth]{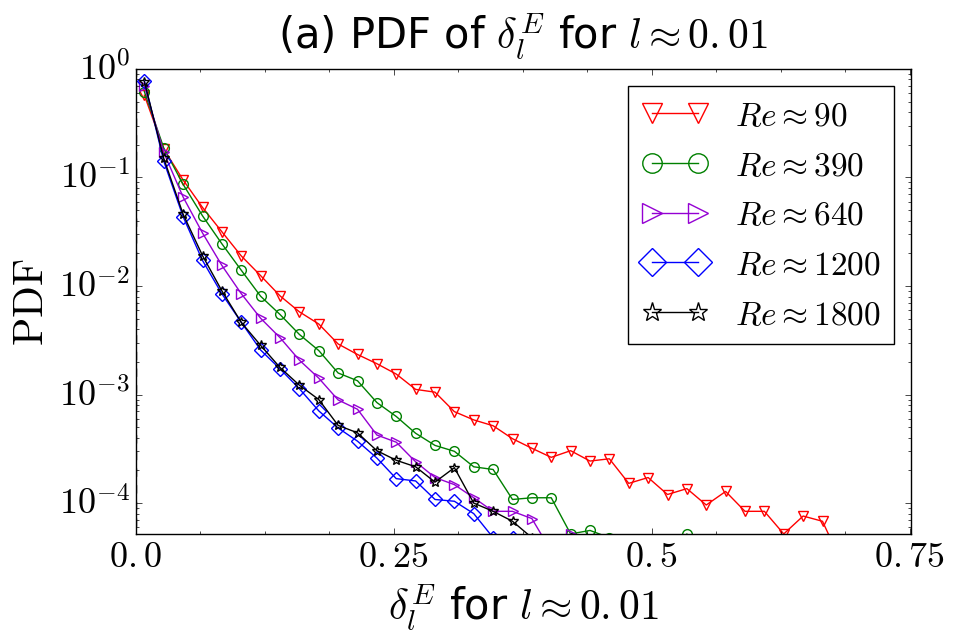}
\hfill
\includegraphics[width=0.49\textwidth]{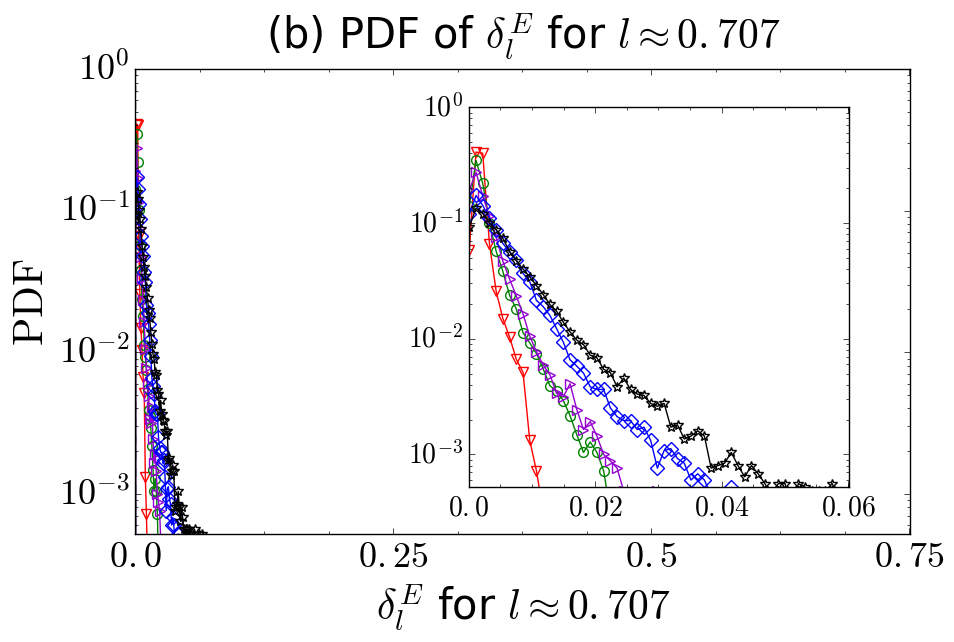}
\vspace*{0.3cm}
\includegraphics[width=0.49\textwidth]{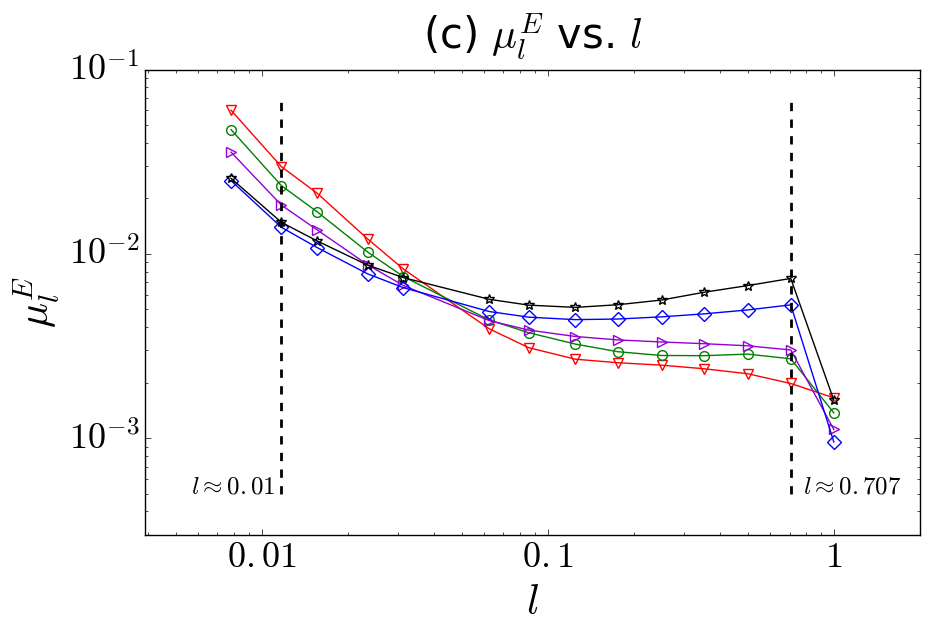}
\hfill
\includegraphics[width=0.49\textwidth]{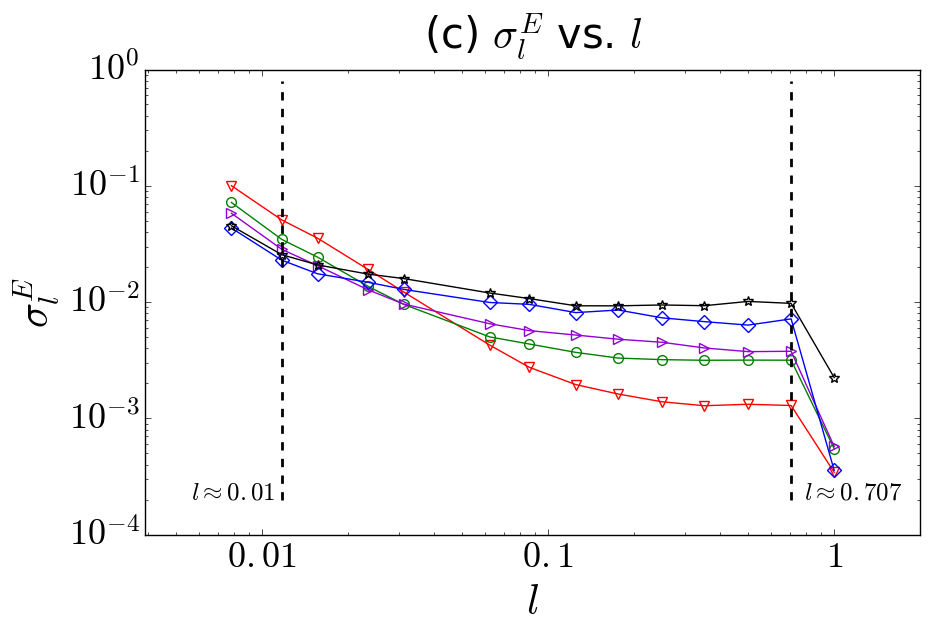}
\caption{Statistics of the balancing error obtained from the kinetic energy balance {\color{black}$\delta_l^{E}$} (see Eq.~\eqref{eq:delta_E}) against the {\color{black}size of the sub-volume $l$ for 5 forced LBGK simulation of different Reynolds numbers. Top figures are PDF of the balancing error for sub-volumes corresponding to $l \approx 0.01$ (Panel (a)) and $l \approx 0.707$ (Panel (b))}. Bottom figures are the mean (Panel (c)) and the standard deviation (Panel (d)) of the balancing error.}
\label{fig:results_E_LBGK}
\end{figure}

\begin{figure}[H]
\centering
\includegraphics[width=0.49\textwidth]{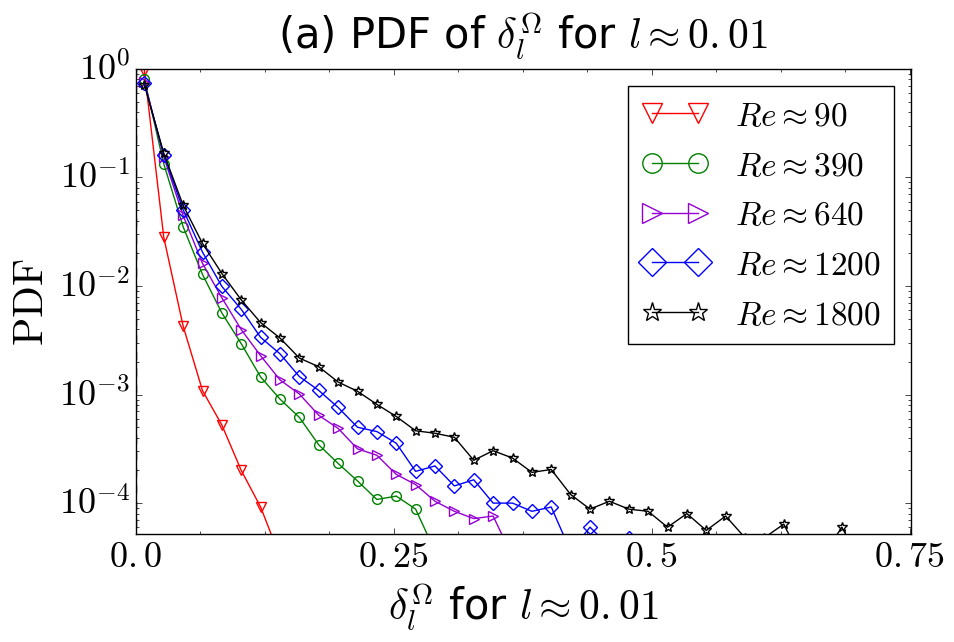}
\hfill
\includegraphics[width=0.49\textwidth]{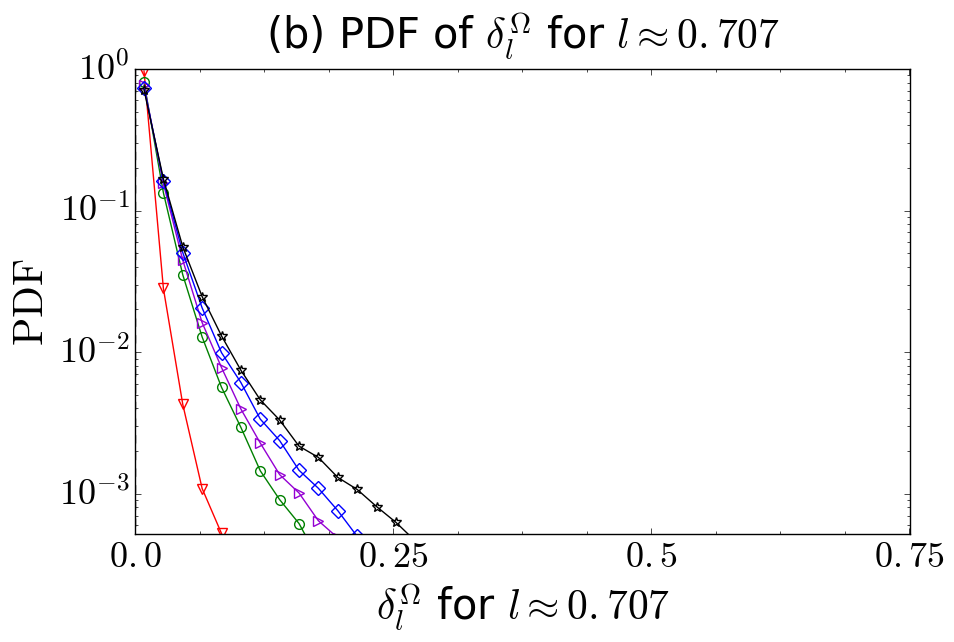}
\vspace*{0.3cm}
\includegraphics[width=0.49\textwidth]{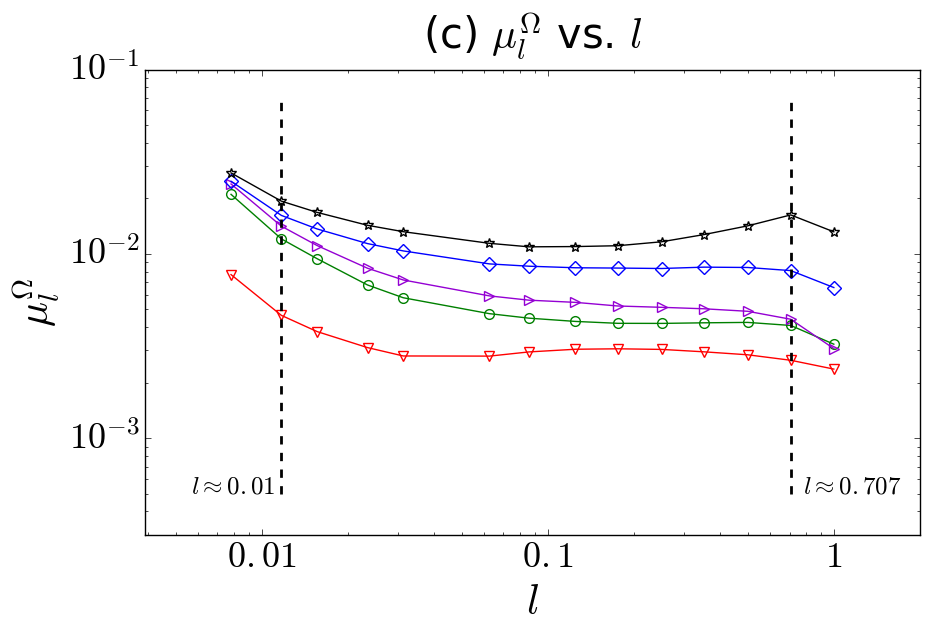}
\hfill
\includegraphics[width=0.49\textwidth]{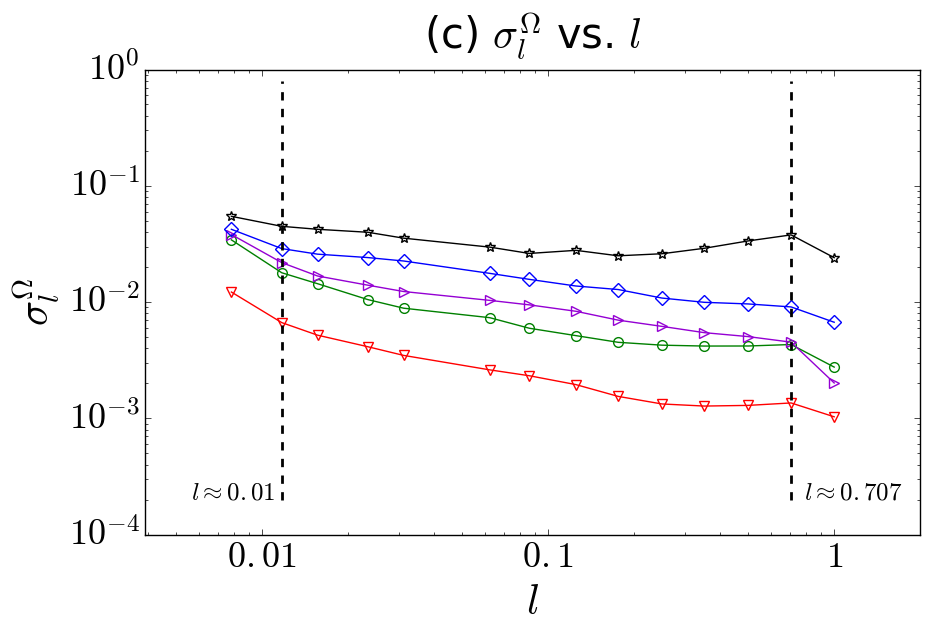}
\caption{Statistics of the balancing error obtained from the enstrophy balance {\color{black} $\delta_l^{\Omega}$} (see Eq.~\eqref{eq:delta_Z}) against the {\color{black} size of the sub-volume $l$ shown for 5 forced LBGK simulation of different Reynolds numbers. Top figures are PDF of the balancing error for sub-volumes corresponding to $l \approx 0.01$ (Panel (a)) and $l \approx 0.707$ (Panel (b))}. Bottom figures are the mean (Panel (c)) and the standard deviation (Panel (d)) of the balancing error.}
\label{fig:results_Z_LBGK}
\end{figure}

\begin{figure}[H]
\centering
\includegraphics[width=0.49\textwidth]{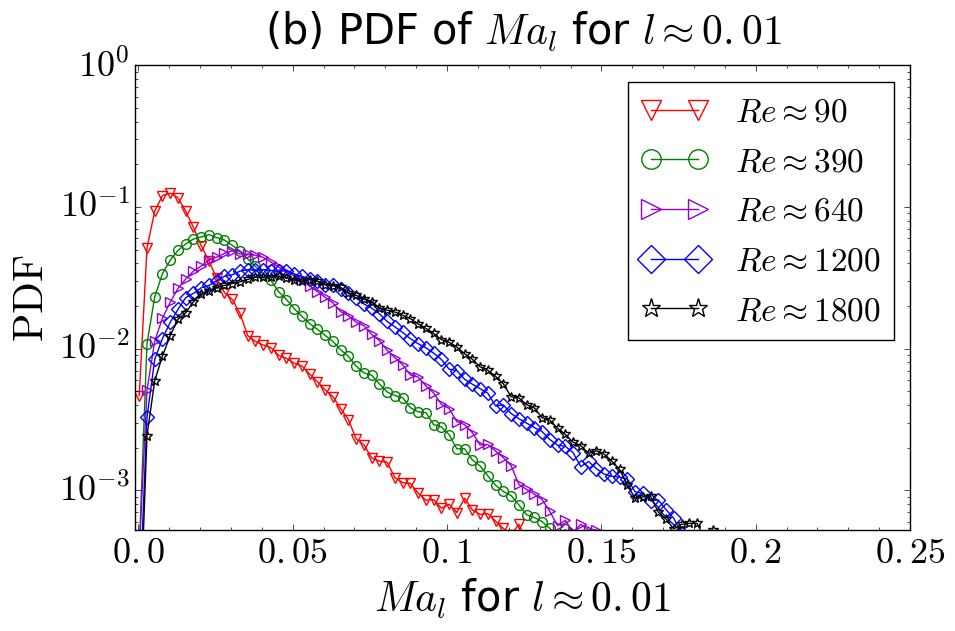}
\hfill
\includegraphics[width=0.49\textwidth]{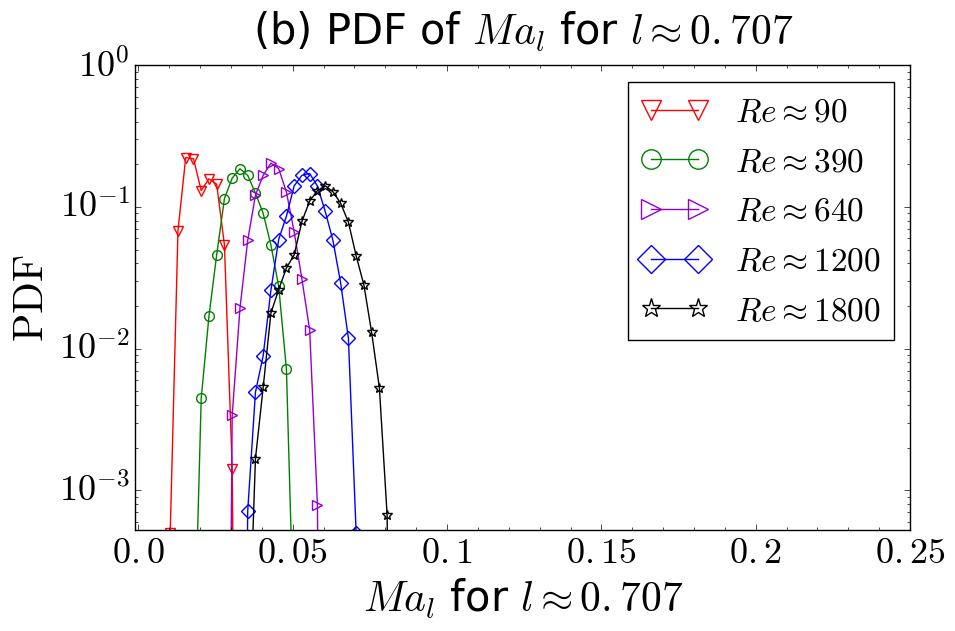}
\vspace*{0.3cm}
\includegraphics[width=0.49\textwidth]{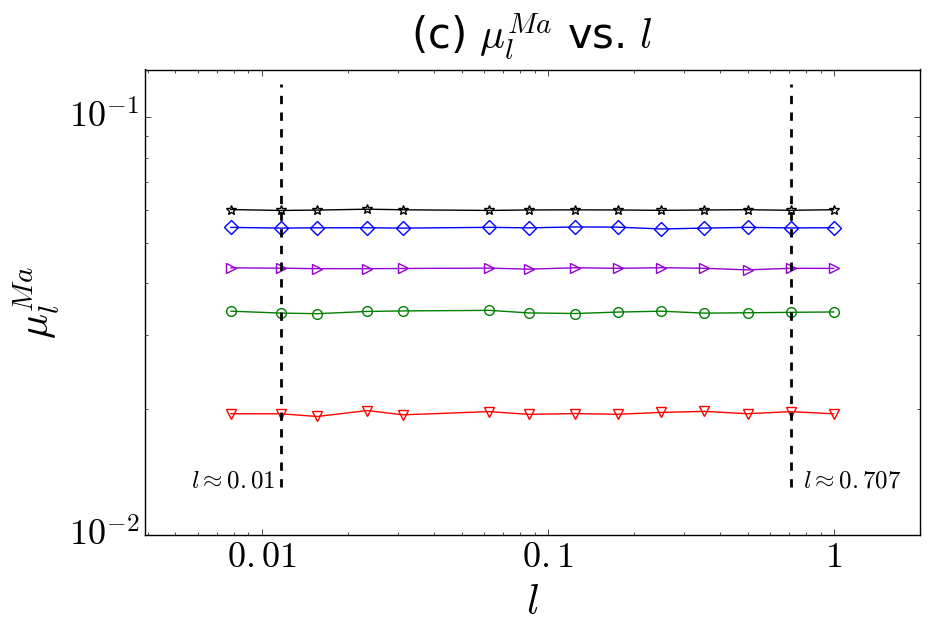}
\hfill
\includegraphics[width=0.49\textwidth]{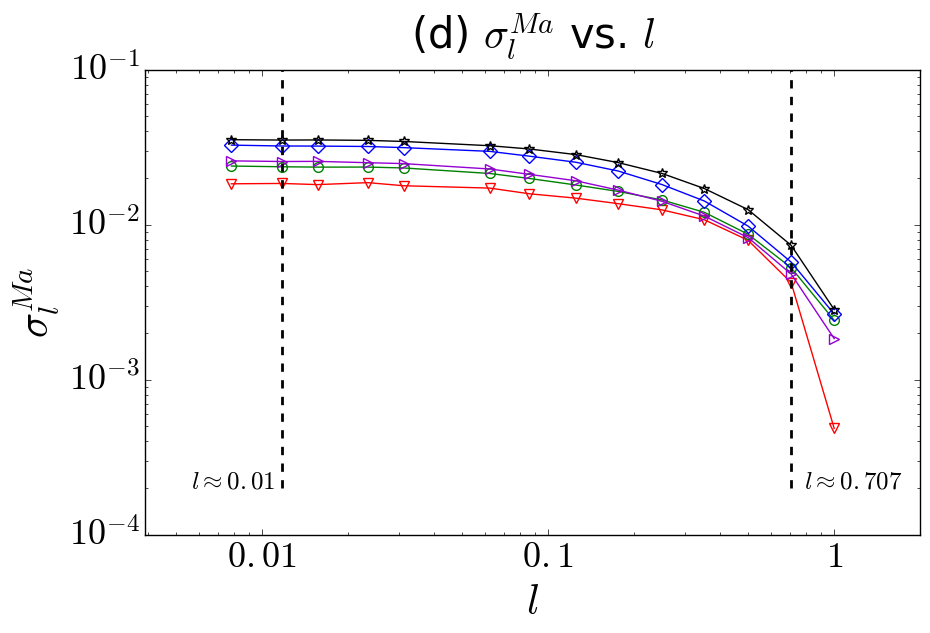}
\caption{Statistics of the Mach number at {\color{black} $Ma_l$ normalized sub-volume size $l$ (see Eq.~\eqref{eq:MaL}) $Ma_l$ against the normalized size of the sub-volume $l$ shown for 5 forced LBGK simulation of different Reynolds numbers. Top figures are PDF of the balancing error for sub-volumes corresponding to $l \approx 0.01$ (Panel (a)) and $l \approx 0.707$ (Panel (b)). Bottom figures are the mean (Panel (c)) and the standard deviation (Panel (d)) of $Ma_l$}.}
\label{fig:mach_LBGK}
\end{figure}

Setting up the forcings as described in section \ref{sec:3}, {\color{black} we analyze configurations of statistically stationary simulations at five different Reynolds numbers $Re \approx 90$, $390$, $640$, $1200$ and $1800$ respectively corresponding to relaxation times $\tau_0 = 0.60$, $0.54$, $0.53$, $0.52$ and $\tau_0^{last}=0.515$, beyond which LBGK is no longer stable}. We then obtain statistics of the balancing errors by averaging both in space and in time on 25 different configurations (see Fig.~\ref{fig:sampling_LBGK}). We show in Fig.~\ref{fig:spectra_LBGK} the superposed time-averaged spectrum for the conducted simulations. At large scales, we can see the effect of the energy removal preventing the energy to accumulate and maintaining the large-scale slope over the backward energy cascade slope of $-\frac{5}{3}$. On the other hand, at small scales, we observe that when we decrease $\tau_0$ ({\color{black} that is, increasing $Re$}) the flow becomes more turbulent and the slope gets increasingly closer to the forward enstrophy cascade slope of $-3$~\cite{Boffetta2012, Frisch1995}.
We present the results of the statistical analysis of the kinetic energy balancing error {\color{black}$\delta_l^{E}$} and the enstrophy balancing error {\color{black} $\delta_l^{\Omega}$} in Figs.~\ref{fig:results_E_LBGK} and~\ref{fig:results_Z_LBGK} respectively. As expected from the LBGK-PS validation results, {\color{black}the hydrodynamic recovery largely depends on the size of the sub-volume it is measured on. Indeed, hydrodynamic recovery is again excellent on large sub-volumes with an order of magnitude of up to ${\cal O} (10^{-3})$, than on small sub-volumes, where we obtain an error that is of orders of magnitude ${\cal O} (10^{-1})$ (see dashed lines in Figs.~\ref{fig:results_E_LBGK} and~\ref{fig:results_Z_LBGK}, Panels (c)-(d)). For the energy balancing error presented Fig.~\ref{fig:results_E_LBGK}, we observe a small dependence on the Reynolds number. However, as shown on Fig.~\ref{fig:results_Z_LBGK}, the enstrophy balance becomes better by decreasing Reynolds number, as it is expected for a quantity that is strongly sensitive to the small-scales resolution.}\\
Having forced with fixed forcing amplitudes, the Mach number of the conducted simulations also varies as a {\color{black} function of the Reynolds number}. To highlight potential high Mach number effects, we plot again the statistics on the Mach number at {\color{black} sub-volume size $l$, $Ma_l$ (Eq.~\refeq{eq:MaL})} as shown in Fig.~\ref{fig:mach_LBGK}. We observe that we are working with Mach number that are qualitatively and quantitatively similar to the ones studied in the previous section (see Fig.~\ref{fig:mach_decaying_LBGK}), hence we conclude again that we work on a range of Mach number that does not impact the hydrodynamics.

\section{Concluding remarks}\label{sec:6}
  We have proposed a general tool to check the generated hydrodynamics of fluid flow simulations. The tool hinges on the calculation of the kinetic energy and the enstrophy balance equation terms averaged over randomly chosen {\color{black} sub-volumes of different size}. We have defined balancing errors, representing the accuracy of the hydrodynamic recovery across {\color{black} sub-volume sizes} and conducted a statistical analysis in the context of 2D homogeneous isotropic turbulence. Firstly, we validated this tool on decaying 2D turbulence by systematically comparing an ensemble of LBGK simulations with an ensemble of PS simulations, both initialized with the same configurations. {\color{black} The PS simulations hydrodynamic recovery accuracy is two to six orders of magnitudes higher than the LBGK simulations'. Moreover, in all cases hydrodynamic recovery is better verified by looking at larger and larger sub-volumes}. {\color{black}Besides, although the enstrophy balance involves higher order derivatives than those present in the kinetic energy equation~\cite{Biferale2010}, the associated extra discretization error  was shown to be negligible as both statistics of the energy and enstrophy balancing errors shows similar order of magnitudes}. Secondly, we have applied this tool to check LBGK hydrodynamic in the context of forced 2D turbulence at increasing Reynolds number. All in all, we have observed statistics of the balancing errors both from kinetic energy balance and enstrophy balance that are very similar to the validation LBGK ensemble's results. In both the validation and benchmark, the Mach number was maintained low enough for its effect to be sub-leading in the hydrodynamic recovery. \\
  The ideal continuation of this work is the study of hydrodynamic recovery with LBM in presence of SGS models of eddy viscosity. To this aim, the developed tool is particularly useful, since it allows to quantitatively describe the effects of under-resolution and the possible improvements led by the SGS model. {\color{black} An expansion of this tool to 3D turbulence is also being developed. Indeed, 3D turbulence is of interest, as it exhibits a direct cascade of energy with a Kolmogorov-predicted slope of $k^{\frac{5}{3}}$, which does not ensure that the flow remains differentiable.}

\section*{Acknowledgement}

The authors would like to thank Fabio Bonaccorso and Michele Buzzicotti at the University of Rome ``Tor Vergata'' for their support in conducting the PS simulations. This work was supported by the European Unions Framework Programme for Research and Innovation Horizon 2020 (2014-2020) under the Marie Sk\l{}odowska-Curie grant [grant number 642069] {\color{black} for the High Performance Computing in Life sciences, Engineering and Physics (HPC-LEAP) project} and by the European Research Council under the ERC grant [grant number 339032]. It is also part of the research programme CSER [project number 12CS034], which is (partly) financed by the Netherlands Organisation for Scientific Research (NWO).

\section*{References}
\bibliographystyle{unsrt}
\bibliography{references}

\end{document}